\documentclass[useAMS,usenatbib]{mn2e}

\usepackage{graphicx}
\usepackage{epsfig}
\usepackage{multicol}

\title[Magnetic field strength in the torus of AGN]{Estimations of the magnetic field strength in the torus of AGN using near-infrared polarimetry.}
\author[E. Lopez-Rodriguez et al.]{E. Lopez-Rodriguez$^{1,2}$\thanks{E-mail:
elr@astro.ufl.edu}, C. Packham$^{1,2}$, S. Young$^{3}$, M. Elitzur$^{4}$, N. A. Levenson$^{5}$, \newauthor
R. E. Mason$^{6}$, C. Ramos Almeida$^{7,8}$, A. Alonso-Herrero$^{9}$\thanks{Augusto Gonzalez Linares Senior Research Fellow}, T.~J.~Jones$^{10}$,\newauthor
 E. Perlman$^{11}$\\
$^{1}$Department of Physics \& Astronomy, University of Texas at San Antonio, One UTSA Circle, San Antonio, TX 78249, USA.\\
$^{2}$Department of Astronomy, University of Florida, 211 Bryant Space Science Center, P.O. Box 11205, Gainesville, FL 32611-2055, USA.\\
$^{3}$Centre for Astrophysics Research, Science \& Technology Research Institute, University of Hertfordshire, Hatfield, AL10 9AB, UK.\\
$^{4}$Department of Physics and Astronomy, University of Kentucky, Lexington, KY 40506, USA.\\
$^{5}$Gemini Observatory, Southern Operations Center, c/o AURA, Casilla 603, La Serena, Chile.\\
$^{6}$Gemini Observatory, Northern Operations Center, 670 N. A'ohoku Place, Hilo, HI 96720.\\
$^{7}$Instituto de Astrof\'isica de Canarias, C/ V\'ia L\'actea, s/n, E-38205, La Laguna, Tenerife, Spain.\\
$^{8}$Departamento de Astrof\'isica, Universidad de La Laguna, E-38205, La Laguna, Tenerife, Spain.\\
$^{9}$Instituto de F\'isica de Cantabria, CSIC-Universidad de Cantabria, 39005 Santander, Spain.\\
$^{10}$Department of Astronomy, University of Minnesota, 116 Church Street S.E., Minneapolis, MN 55455, USA.\\
$^{11}$Department of Physics and Space Sciences, 150 W. University Blvd., Florida Institute of Technology, Melbourne, FL 32901, USA}

\newcommand{\degree}{\ensuremath{^\circ}}
\newcommand{\um}{\ensuremath{\umu}m}
\newcommand{\K}{K$_{\rm n}$}
\newcommand{\Av}{A$_{\rm v}$}

\begin{document}

\date{Accepted by MNRAS. Accepted 2013 February 25. Received 2013 February 11; in original form 2012 December 05}

\pagerange{\pageref{firstpage}--\pageref{lastpage}} \pubyear{2012}

\maketitle

\label{firstpage}


\begin{abstract}

An optically and geometrically thick torus obscures the central engine of Active Galactic Nuclei (AGN) from some lines of sight. From a magnetohydrodynamical framework, the torus can be considered to be a particular region of clouds surrounding the central engine where the clouds are dusty and optically thick. In this framework, the magnetic field plays an important role in the creation, morphology and evolution of the torus. If the dust grains within the clouds are assumed to be aligned by paramagnetic alignment, then the ratio of the intrinsic polarisation and visual extinction, P(\%)/\Av, is a function of the magnetic field strength.

To estimate the visual extinction through the torus and constrain the polarisation mechanisms in the nucleus of AGN, we developed a polarisation model to fit both the total and polarised flux in a 1.2$''$ ($\sim$ 263 pc) aperture of the type 2 AGN, IC5063. The polarisation model is consistent with the nuclear polarisation observed at \K~(2.0 - 2.3 \um) being produced by dichroic absorption from aligned dust grains with a visual extinction through the torus of 48$\pm$2 mag. We estimated the intrinsic polarisation arising from dichroic absorption to be P$^{\mbox{\sevensize{dic}}}_{\mbox{\sevensize{\K}}} =$ 12.5$\pm$2.7\%.

We consider the physical conditions and environment of the gas and dust for the torus of IC5063. Then, through paramagnetic alignment, we estimate a magnetic field strength in the range of 12 - 128 mG in the NIR emitting regions of the torus of IC5063. Alternatively, we estimate the magnetic field strength in the plane of the sky using the Chandrasekhar-Fermi method. The minimum magnetic field strength in the plane of the sky is estimated to be 13 and 41 mG depending of the conditions within the torus of IC5063. These techniques afford the chance to make a survey of AGN, to investigate the effects of magnetic field strength on the torus, accretion, and interaction to the host galaxy.

\end{abstract}




\section{Introduction}

The detection of polarised broad emission lines in the spectrum of NGC1068 revealed an obscured Active Galactic Nucleus (AGN) through scattering of the broad line region (BLR) radiation \citep*{AM85}. This study gave a major boost to the unified model \citep{A93,UP95} of AGN, which hods that all AGN are essentially the same object, viewed from different line of sight (LOS). In the unified model scheme, the AGN classification solely depends on the obscuration of an optically and geometrically thick dusty torus

Recent studies \citep[e.g.][]{N02,N08a} have proposed that the dust in the torus is distributed in clumps, and accounts for (a) the distribution of temperatures along the torus; and (2) the variety of spectral energy distributions (SED) produced by geometry, clumpy distribution and spectral features, such as the 10 $\um$ absorption and emission. The clumpy torus model holds that the dusty clumps are distributed in a few parsecs, consistent with observations \citep[e.g.][]{R02,J04,P05,S06,T07}. This means that the torus cannot be resolved at optical/infared (IR) wavelengths, even with the high-spatial resolution provided by 8-m class telescopes. Recent high-spatial resolution observations of Seyfert galaxies have shown that the clumpy torus model can account for the near-IR (NIR) and mid-IR (MIR) emission \citep[see][]{M06,RA09,RA11,AH11}. Some studies \citep*{Mor2009,Mor2012} have shown that a component composed by hot graphite dust can explain the NIR emission of Type 1 AGN. The clumpy torus models permit, from a statistical view, an examination of the general properties of the torus, i.e. inclination to our LOS, number of clouds, covering factor, optical depth of clouds, etc. However, the intrinsic properties of the clumpy torus, i.e. the dust grain composition, grain size, grain alignment, etc. remain unknown. Polarimetry techniques show a powerful potential to obtain the intrinsic properties of the torus. Several polarimetric studies in the IR to NGC1068 \citep{M07,P07} have demonstrated this potential, constraining (1) interstellar dust properties; (2) upper-limit diameter of the torus; (3) mechanisms of polarisation  and hence magnetic field directions in the central few pc. 

Several models to explain the existence of the torus have been proposed. Some models explain the torus as an inflow of gas from large scales \citep[e.g.][]{W09, SKB11}. \citet{W09} presented numerical simulations of the interstellar medium to track the formation of molecular hydrogen forming an inhomogeneous disc around the central engine, identified as the torus. \citet*{SKB11} suggested the origin of the torus as clouds falling to the central engine. In this scheme, it is difficult to explain how the energy dissipation of clouds and cloud-cloud collision tends to concentrate the clouds to form the torus near the midplane \citep*{KB88}. Conversely, some models \citep{BP82,EBS92,ES06} suggest that the clouds are confined by the magnetic field generated in the central engine and are accelerated by the hydromagnetic wind. In this scheme, the hydromagnetic wind could lift the clouds from the midplane to form a geometrically thick distribution of clouds surrounding the central engine, where the torus is in a particular region of the hydromagnetic wind where the clouds are dusty and optically thick. \citet{KKE1999} suggests that magnetic field strengths greater than 20 mG can account for a clumpy disc-driven wind model, where clouds moves along the magnetic field lines in a homogeneous outflow component. VLA circular polarimetry observations of NGC4258 inferred an upper-limit of the magnetic field strengths of 300 mG \citep{H1998}. Further polarimetric observations \citep{Modjaz2005} at 22GHz of the water vapour maser clouds in NGC4258 using the VLA and GBT, estimate magnetic field strength from 90 to 300 mG at a radius of 0.2 pc from the central engine. In summary, observations and theoretical models for maser clouds at distances $\sim$0.2 pc seems to be in agreement. However, observational or theoretical studies at larger distances, where the clouds are optically thick and dusty, have not been done to date. 

The nearby (z = 0.01\footnote{Through this work, we adopt H$_{0} =$ 73 km s$^{-1}$ kpc$^{-1}$, so that 1$'' =$ 219 pc for the redshift of IC5063}; \citet{deV1991}) elliptical galaxy (r$^{1/4}$ brightness profile; \citet{C91}) IC5063 (PKS 2048-572) shows polarised scattered H$_{\alpha}$ broad emission lines suggesting a hidden Type 1 AGN \citep{I93}, and an increase in the degree of polarisation in the infrared at J, H and K \citep{H87}. A prominent dust lane has been observed \citep*{C91} along the long-axis of IC5063. IC5063 shows a radio luminosity (log P$_{1.4GHz} =$ 23.8 W Hz$^{-1}$, H$_{0} =$ 50 km s$^{-1}$ kpc$^{-1}$, \citet*{C91}) about two orders of magnitude larger than is typical nearby Seyfert galaxies, classify IC5063 in the range of low-luminosity radio elliptical galaxies. The central polarised source have been suggested to be a BL Lac object from infrared observations by \citet{H87}. They showed a steep-spectrum IR component and suggested a synchrotron central source. Further NIR studies at J, H and K bands by \citet{B90b} modeled the total and polarised flux of IC5063, and concluded that the nuclear source can be explained by a reddened central source with an power-law index of 1.5 (typical for Seyfert 1). 

In this paper we estimate the magnetic field strength in the NIR emitting regions of the torus of AGN, through NIR polarimetric observations of IC5063 in the J, H and \K~bands. We develop a polarisation model to simultaneously fit the total and polarised flux, that allow us to interpret the different mechanisms of polarisation to the central engine of IC5063. Further estimates of the extinction to the central engine at different wavelengths allow us to interpret the origin of the NIR polarisation. Finally, four independent methods of calculating the magnetic field strength in the clumpy torus of IC5063 are presented.

The paper is structured as follow. In Section 2, we describe the observations and data reduction. In Section 3, the results are shown, then analysed in Section 4. We discuss various mechanisms of polarisation that could account for the polarisation in the nucleus of IC5063 and we develop a polarisation model to interpret our data in Section 5. In Section 6, the nuclear extinction estimated at various wavelengths is discussed. In Section 7,  the magnetic field strength of the torus of IC5063 is estimated. Finally, conclusions are presented in Section 8.



\section[]{Observations and data reduction}

The observations were made using the infrared polarimeter built by the University of Hertfordshire \citep{H94} for the Anglo-Australian Telescope (AAT). The polarimeter features consist of an achromatic  (1 - 2.5 \um) half-wave plate (HWP) retarder from 1.0 to 2.5 \um, which can be stepped to set angular positions. This is placed upstream of the observatory's near-infrared camera, IRIS \citep{GL90}, in which a magnesium fluoride Wollaston prism was used as an analyser. A mask in the focal plane of IRIS served to blank out half of the field so that the ordinary (o-ray) and extraordinary (e-ray) ray from Wollaston prism do not overlap when imaging extended objects as well as reduce the sky from overlapping.

IC5063 was observed on 1994 July 25, in the J, H and \K~(2.0 - 2.3 \um)  bands, with IRIS at the f/15 focus using the Rockwell camera with 128 $\times$ 128 pixels. The pixel-scale is 0.6$''$~per pixel, giving a field of view of 76$''$~$\times$~76$''$. The half-wave retarder was stepped sequentially to four position angles (PA; 0\degree, 45\degree, 22\fdg5 and 67\fdg5) taking an exposure at each HWP PA. Images at each HWP PA were flat-fielded, sky-subtracted and then cleaned by interpolating over dead and hot pixels and cosmic rays. Next, the images were registered and shifted by fractional pixel amounts in order to account for slight image drift between frames. Then, the polarisation images were constructed using our own IDL routines. The individual o- and e-ray images were co-averaged to increase the signal-to-noise ratio. Next,  the o- and e-rays were extracted using a rectangular aperture of  28$''$~$\times$~18$''$. The Stokes parameters, I, Q and U, were calculated according to the ratio method prescription (see \citet{Tinbergen1996}). Finally, the degree of polarisation, P, was derived such as:

\[
\mbox{P} =  100 \times \frac{\sqrt{Q^2+U^2}}{\mbox{I}} 
\]

\noindent
and the position angle of polarisation, $\theta$, was found by:

\[
\theta  =  \frac{1}{2}\tan^{-1}{\left(\frac{U}{Q}\right)}
\]

A summary of observations is shown in Table \ref{table1}. The night was photometric with the seeing estimated to be 1.2$^{''}$ from the automatic-guide at the observatory. Flux standard stars were not observed during this night, so we used the photometric data from \citet{A82}. Specifically, the flux calibration was performed using photometry at J, H and K in a 10$''$ aperture, where the large aperture ensures that centering issues are minimal. The difference between the K and \K~bands makes a small contribution to the total photometric error. The zero-flux standard values \citep{CRL1985} of 1603 Jy, 1075 Jy and 667 Jy for J, H and \K, respectively were used to transform between magnitudes and fluxes. To calibrate the PA of polarisation at J, H and \K, a 4.5$''$ aperture from \citet{H87} was used. Since unpolarised standard stars were unavailable, an instrumental polarisation of 0.02$\pm$0.06\%, from previous studies by \citet{H94} was adopted. 


 \begin{table}
 \centering
 \begin{minipage}{80mm}
  \caption{Summary of observations.}
  \label{table1}
  \begin{tabular}{@{}cccc@{}}
  \hline
 {\bf  Filter}       & {\bf Frame Time}     & {\bf Sets}  &  {\bf Total observation time} \\
                         &  {\bf (s)  }           &           &          {\bf  (s)  }               \\
 \hline
     J               &    30              & 12    &      1440                    \\
    H               &    20              & 8      &      640                       \\
   \K                &   10              & 8      &       320                       \\
\hline
\end{tabular}
\end{minipage}
\end{table}


 

\section{Results}

\subsection{Photometry}

\begin{figure}
 \centering
 \begin{minipage}{85mm}
 \includegraphics[angle=0,scale=0.5,clip,trim=2.5cm 14cm 2cm 3cm]{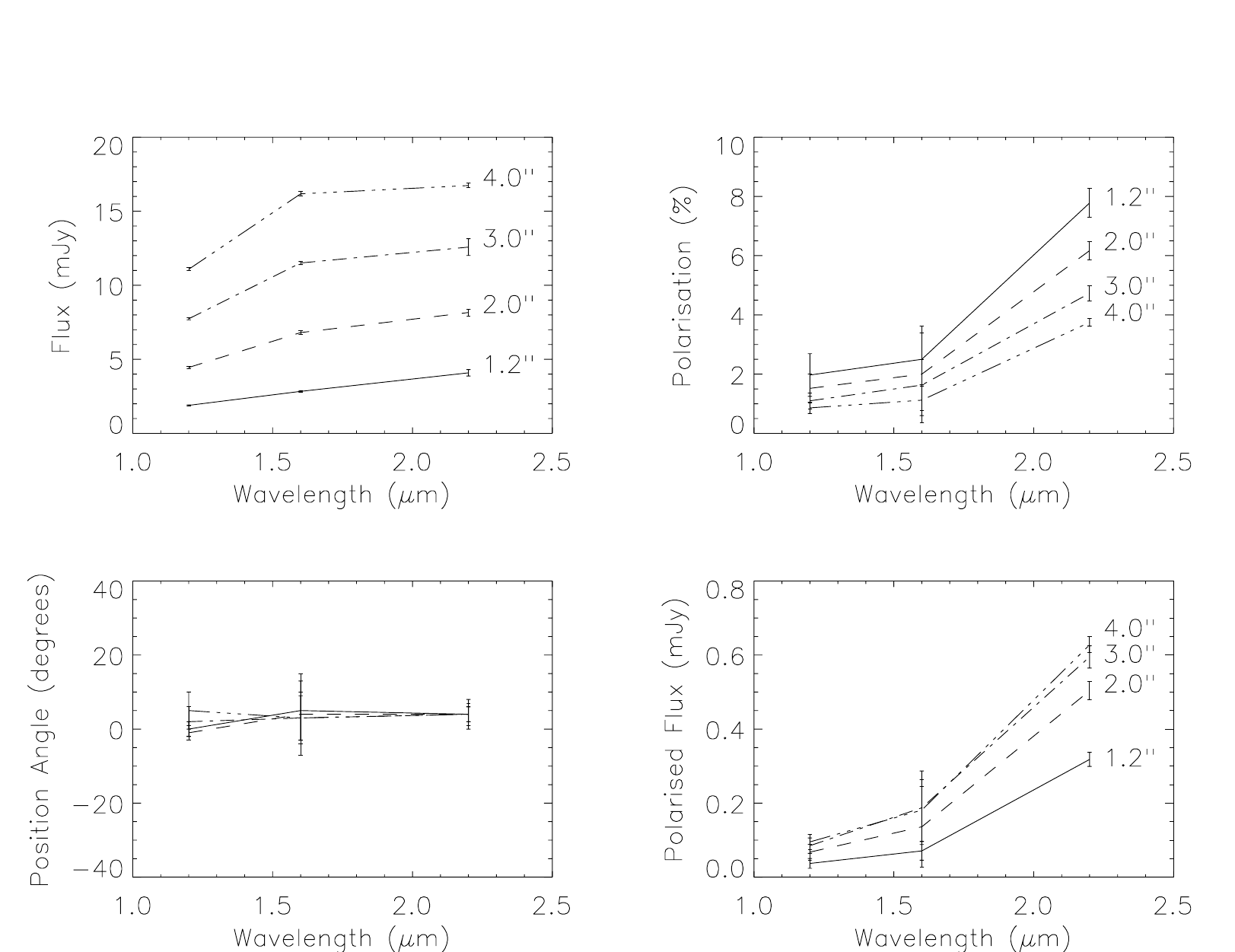}
\caption{Total flux ({\it top left}), degree ({\it top right}) and position angle ({\it bottom left}) of polarisation and polarised flux ({\it bottom right}) versus wavelength. Measurements with aperture size of 1.2$''$ ({\it solid line}), 2.0$''$ ({\it dashed line}), 3.0$''$ ({\it dashed dotted line}) and 4.0$''$ ({\it dashed three-dotted line}) are shown in each plot.}
\label{fig1}
\end{minipage}
\end{figure}

We made measurements of the nuclear total flux in several apertures to compare to previously published values (Table \ref{table2}). In all cases, photometric errors were estimated by the variation of the counts in subsets of the data.  The difference in fluxes between our and previously published results is $<$ 4\% in all apertures. To optimally measure the AGN flux of IC5063 and minimize the contributions from the host galaxy, dust lane and ionisation cones, photometry in an aperture equal to the seeing of the observations (1.2$''$) was used.  To investigate the flux dependence on the aperture size, photometry in three apertures, 2.0$''$, 3.0$''$ and 4.0$''$, was measured (Table \ref{table3}). An increase in total flux at both longer wavelengths and larger aperture was found (Figure \ref{fig1}). In line with previous studies, an elliptical profile with a FWHM of 3.7$''$~$\times$~2.6$''$, 3.4$''$~$\times$~2.5$''$ and 2.3$''$~$\times$~1.8$''$ at J, H and \K~respectively, was found for the nucleus of IC5063. In all filters, IC5063 was found to be extended along PA $=$ 75\degree. The FWHM was estimated using a Gaussian profile. 


 \begin{table*}
  \caption{Comparison with literature of the nuclear photometry of IC5063.}
  \label{table2}
  \begin{tabular}{cccccc}
  \hline
  {\bf Aperture}     &   {\bf Filter }		& \multicolumn{2}{c}{\bf  \citet{H87}  }                   &     \multicolumn{2}{c}{\bf This work}    	\\
      {}                      &         {}                     &  {\bf Mag}  		 & {\bf P}  			&  {\bf Mag}  			& {\bf P}    \\
      {\bf   ($''$) }     &         {}                     &  {\bf (mag)}              & {\bf (\%)} 		&   {\bf (mag)}              	& {\bf (\%)} 		\\
      \hline
        2.25      &                 H                     &      12.38            & 1.75$\pm$0.34		&    12.86 $\pm$ 0.02	& 1.8$\pm$1.5	               \\
                      &		    \K			&	11.79            & 6.33$\pm$0.50		&   12.18  $\pm$ 0.02	& 6.0$\pm$0.3 	        \\
       4.5	   &		     J			&	12.50	   & 0.77$\pm$0.07		&    12.71  $\pm$ 0.01	& 0.7$\pm$0.2		        \\
       		   &	             H			&       11.80	   & 0.99$\pm$0.05		&    11.83   $\pm$ 0.01	& 1.0$\pm$0.4		\\
		   &		   \K			&	11.10	   & 3.25$\pm$0.45		&    11.34  $\pm$ 0.01	& 3.3$\pm$0.1      	\\
\hline
{\bf Aperture  }   &		{\bf Filter }		&   {\bf \citet{A82}  }    &  {\bf  This work }       \\
     {}                     &         {}                                &  {\bf Mag}  & {\bf Mag} \\
     {\bf   ($''$)  }    &                                         &  {\bf (mag) }  & {\bf (mag)  }    \\
 \hline
       5.0	   &		     J			&	12.72	   &  12.59	$\pm$ 0.01	        \\
       		   &	             H			&       11.80	   &  11.72 $\pm$ 0.01		\\
		   &		   \K			&	11.32	   &  11.23 $\pm$ 0.01		\\		                  
\hline
\end{tabular}
\end{table*}



 \begin{table}
 \centering
 \begin{minipage}{80mm}
\caption{The observed degree and position angle of polarisation and photometry of IC5063.}
\label{table3}
  \begin{tabular}{@{}ccccc@{}}  
  \hline
{\bf Aperture}   & {\bf Filter} &{\bf P} & {\bf $\theta$} & {\bf Magnitude }  \\
{\bf ($''$)}  &         & {\bf (\%)}   & {\bf (\degree)} &   \\
\hline
1.2 & J   &   2.0 $\pm$ 0.7  &  0 $\pm$ 2 & 14.81 $\pm$ 0.02\\
        & H  &   2.5 $\pm$ 0.9  &  5 $\pm$ 8  & 13.90 $\pm$ 0.02\\
         & \K  &  7.8 $\pm$ 0.5  &  4 $\pm$ 4 & 13.01 $\pm$ 0.06\\
2.0 & J   &   1.5 $\pm$ 0.5  &  -1  $\pm$ 2 & 13.88 $\pm$ 0.02\\
        & H  &   2.0 $\pm$ 1.6  &  4 $\pm$ 11 & 12.95 $\pm$ 0.02\\
         & \K  &  6.2 $\pm$ 0.3  &  4 $\pm$ 3 & 12.26 $\pm$ 0.03\\
3.0 & J   &    1.1 $\pm$ 0.3  &  2 $\pm$ 4 & 13.28 $\pm$ 0.01\\
        & H  &    1.6 $\pm$ 0.9  & 3 $\pm$  7   & 12.38 $\pm$ 0.01\\
         & \K  &   4.7 $\pm$ 0.3  &  4 $\pm$ 2 & 11.79 $\pm$ 0.02\\
4.0 & J   &    0.9 $\pm$ 0.2  &  5 $\pm$ 5 & 12.89 $\pm$ 0.01\\
        & H  &    1.1 $\pm$ 0.5  &  3 $\pm$ 6   &  12.01 $\pm$ 0.01\\
         & \K  &   3.8 $\pm$ 0.1  &  4 $\pm$ 2 &  11.48  $\pm$ 0.01\\
\hline
\end{tabular}
\end{minipage}
\end{table} 



\subsection{Polarimetry}

\begin{figure*}
\includegraphics[angle=0,scale=0.66,clip,trim=1.5cm 11cm 2.5cm 1cm]{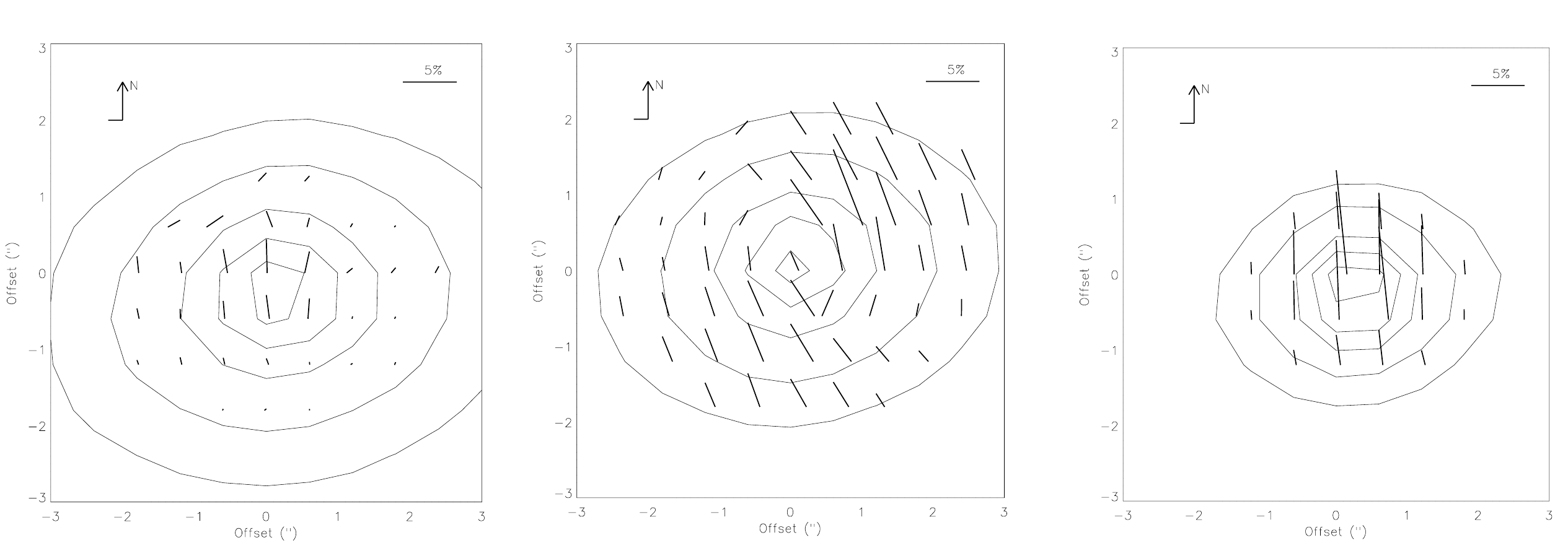}
\caption{Total flux contours with polarisation vectors of the central 6$''$~$\times$~6$''$ at J ({\it left}) , H ({\it middle}) and \K~({\it right}). A vector of 5\% of polarisation is shown at the top right of each plot. The lowest-level total flux contour represents an uncertainty of 0.8\% in the degree of polarisation. Contours are plotted in intervals of 10\% to the peak of the total flux at each band. North is up and East left. The physical scale is 1$"$ = 219 pc.}
\label{fig2}
\end{figure*}

\begin{figure*}
\includegraphics[angle=0,scale=0.28,clip,trim=3cm 16cm 2cm 2cm]{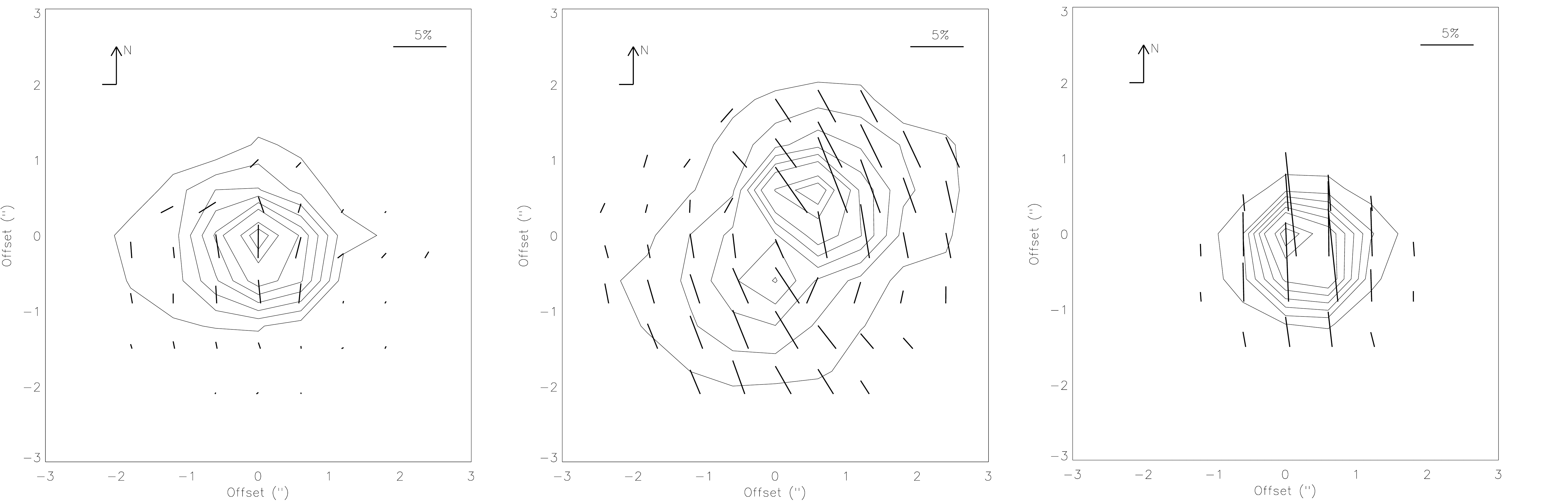}
\caption{Polarised flux contours with polarisation vectors of the central 6$''$~$\times$~6$''$ at J ({\it left}) , H ({\it middle}) and \K~({\it right}). A vector of 5\% of polarisation is shown at the top right of each plot. Contours are plotted in intervals of 10\% to the peak of the polarised flux at each band. North is up and East left. The physical scale is 1$"$ = 219 pc.}
\label{fig3}
\end{figure*}

\begin{figure*}
\includegraphics[angle=0,scale=1.0,clip,trim=2.3cm 16.4cm 2cm 6cm]{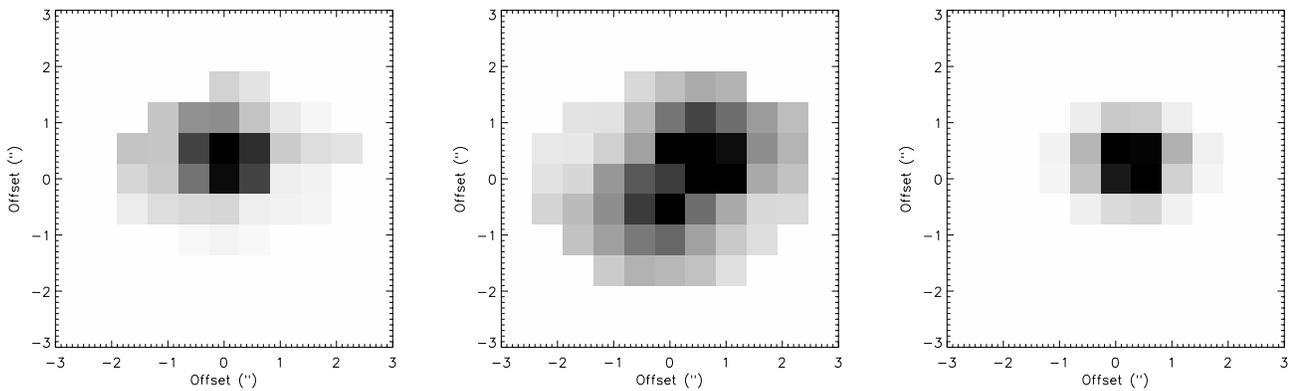}
\caption{Polarised flux images of the central 6$''$~$\times$ 6$''$ at J ({\it left}) , H ({\it middle}) and \K~({\it right}) bands. North is up and East is left. The physical scale is 1$"$ = 219 pc.}
\label{fig4}
\end{figure*}

We made measurements of the nuclear degree of polarisation in several apertures to compare to previously published values (Table \ref{table2}). In all cases, polarimetric errors were estimated by the variation of the counts in subsets of the data The observed degree and PA of polarisation in four different sized apertures, 1.2$''$, 2.0$''$, 3.0$''$, 4.0$''$, were measured (Table \ref{table3}). Polarimetry maps of the total and polarised flux are shown in Figures \ref{fig2} and \ref{fig3} in J, H and \K~bands. In these figures, the overlaid polarisation vectors are proportional in length to the degree of polarisation with their orientation showing the PA of polarisation. The lowest-level total flux contour indicates the level at which 0.8\% of uncertainty in polarisation is reached. The polarised flux images at J, H and \K~are shown in Figure \ref{fig4}. 


\begin{figure}
 \centering
 \begin{minipage}{85mm}
\includegraphics[angle=180,scale=0.48,trim=0cm 2cm 5cm 2cm]{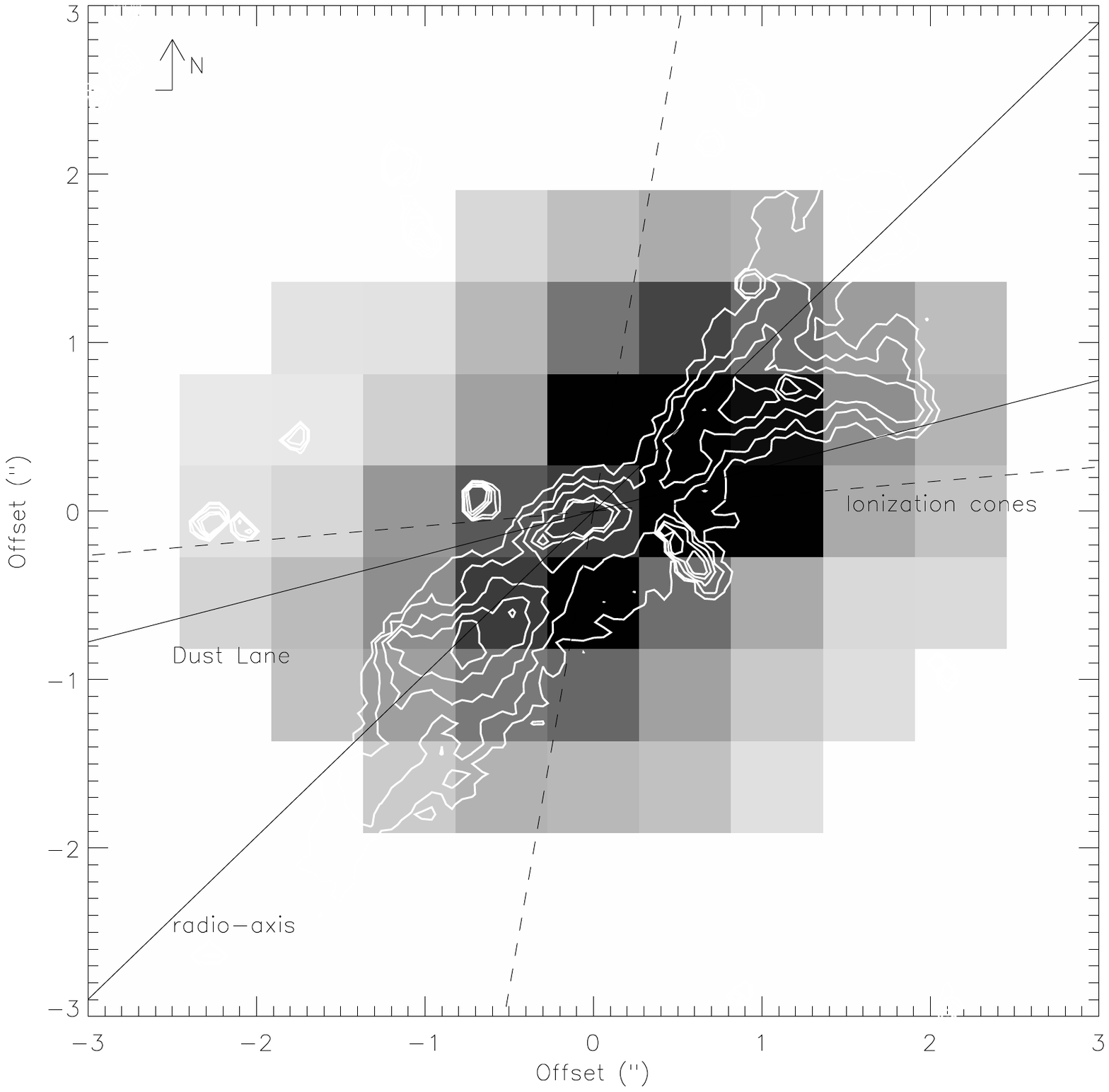}
\caption{Polarised flux at H ({\it grey scale}). [OIII] ionisation structure from HST by \citet{S03} ({\it contours}). Radio-axis with PA$\sim$295\degree is shown \citep{MOT98}. Dust lane axis with PA $=$ 75\degree, from \citet*{C91}. Polarised cones axis ({\it dashed line}) with opening angle of 75\degree$\pm$5\degree are from this paper. The physical scale is 1$"$ = 219 pc.}
\label{fig5}
\end{minipage}
\end{figure}

At J, we interpret the extended polarisation in the central regions to be dichroic polarisation of starlight through aligned dust grains in the host galaxy and the central dust lane \citep*{C91}.  The patchy distribution of dust in the nuclear regions results in a polarisation dependence with aperture size. At H, the most striking feature is the polarisation vector pattern (Figures \ref{fig2} and \ref{fig3}) and the extended structure seen in polarised flux (Figure \ref{fig4}). The polarised structure is extended $\sim$3$''$ from NW to SE along of a PA $\sim$ 300\degree. The strong spatial correspondence with the structure at 8GHz of a PA of 295\degree~\citep*{MOT98}, and the ``X-shape" in the [OIII] observations at PA of 290\degree~\citep*{C91} are shown in Figure \ref{fig5}. The polarised flux images show a tentative opening angle of $\sim$75\degree~at the H band, larger than the 60\degree~measured through the [OIII] ionised structure by \citet*{C91}. We note that other AGN show a wider opening angle in polarised flux than emission line imaging, e.g. NGC1068 has an opening angle of 60\degree$\pm$20\degree~from [OIII] imaging \citep{Evans91} and 80\degree~from NIR polarimetry imaging \citep{P97}. The opening angle in polarised light is larger because of scattering can occur from any material in the medium, instead, the medium needs to be filled by [OII], in order to be ionised ([OIII]), producing a smaller opening angle. We interpreted the biconical polarised distribution to be scattering of light from the central engine to our LOS. At \K, a highly polarised point-like source is observed. 


The polarisation of the central 1.2$''$ of IC5063 at J, H and \K~was measured to be 2.0$\pm$0.7\%, 2.5$\pm$0.9\% and 7.8$\pm$0.5\%, respectively. Figure \ref{fig1} shows that the degree of polarisation decreases as the aperture size increases in all filters, whereas the polarised flux remains similar in the 3$''$ and 4$''$  apertures at J and H. Finally, the PA of polarisation is wavelength-independent (within the error bars) and measured to be 3$\pm$6\degree~in the three filters. This result is consistent with the PA of polarisation of 3$\pm$2\degree~measured by \citet{H87} in the J, H and K filters, and $\sim$3\degree~using optical (0.45 - 0.70 \um) spectropolarimetry by \citet{I93}. Note that the PA of polarisation, 3$\pm$6\degree, is perpendicular to (1) the long-axis, 300\degree, of the galaxy \citep{C91}; and (2) the radio-axis, 295\degree \citep*{MOT98}.


\section{ANALYSIS}

The measured polarisation is the diluted polarised light from the AGN produced by the diffuse stellar emission. Hence, the intrinsic polarisation can be estimated accounting for the contributions by the AGN and diffuse stellar emission to the total flux in the nucleus of IC5063. We quantify these contributions through an examination of the nuclear profiles in Section \ref{nucleartotalflux}.

\subsection{The nuclear total flux}
\label{nucleartotalflux}

\begin{figure}
 \centering
 \begin{minipage}{85mm}
\includegraphics[angle=0,scale=0.4,trim=2cm 10cm 9cm 3.5cm]{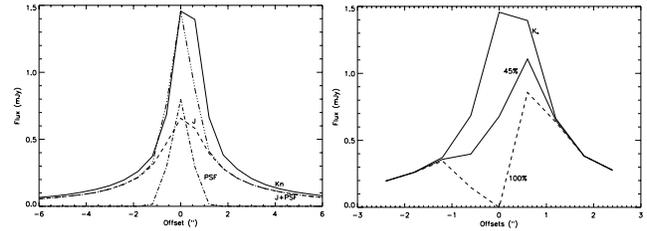}
\caption{Cuts profile for the first method in Section \ref{nucleartotalflux} are shown ({\it left panel}). Cuts profiles at J ({\it dashed line}), \K~({\it solid line}) and the scaled AGN$_{\mbox{\sevensize PSF}}$ ({\it dashed dotted line}). The combined J profile and AGN$_{\mbox{\sevensize PSF}}$ matched with the \K-profile ({\it thick dashed  three-dotted line}). $``$Flat-top" profile for the second method in Section \ref{nucleartotalflux} are shown ({\it right panel}). Same \K~profile than in left panel ({\it solid line}) and the subtraction of the AGN$_{\mbox{\sevensize PSF}}$-scaled at 45\% ({\it thick solid line}) and 100\% ({\it dashed line}).} 
\label{fig6}
\end{minipage}
\end{figure}

The AGN is embedded within diffuse stellar emission in the nuclear bulge. The emission within the nuclear regions of IC5063 can be considered to arise from two dominant emission components: (1) diffuse stellar emission in the nuclear bulge; and (2) emission from the AGN. Photometric cuts through the nucleus in the J band show no evidence of a nuclear point-source, and hence we assume that there is negligible emission from the AGN in our smallest aperture (1.2$''$; $\sim$263 pc). Thus, the J band profile is assumed to be representative of the profile of the diffuse stellar emission in the nuclear bulge. However, at \K~we detect both a central point-source and emission from the diffuse stellar bulge. We interpret the central point-source emission to be the AGN. To estimate the relative contributions from these two emission components, two different methods were followed. In the first method, we followed a similar analysis to that of \citet{T92} and \citet{P96}.  We took photometric cuts at J and \K~along the dust lane axis, PA $=$ 75\degree~\citep*{C91}. These cuts are assumed to be along a line of constant extinction through the nucleus (A$_{\mbox{v}} =$~0.3~mag; \citet*{C91}). The AGN contribution, modeled as a point-source convolved with a Gaussian profile of FWHM equal to that of the seeing disc, is termed AGN$_{\mbox{\sevensize PSF}}$. To fit the \K~nuclear bulge profile along the cut, the emission was modeled as the summation of the AGN$_{\mbox{\sevensize PSF}}$ and the diffuse stellar emission from the J band photometric profile. The photometric profiles in J, \K, the modeled AGN$_{\mbox{\sevensize PSF}}$ and the combined J and AGN$_{\mbox{\sevensize PSF}}$ to fit the \K-profile are shown in Figure \ref{fig6}, showing an acceptable fit to the data. Through this method, the best estimate of the contribution of the AGN emission to the total flux is 40$\pm$2\% in a 1.2$''$ aperture. This result is comparable to that determined for Centaurus A, $\sim$30\% AGN contribution, when using a similar methodology by \citet{P96}.

In the second method, the scaled AGN$_{\mbox{\sevensize PSF}}$ from the first method was used to subtract the AGN emission from the \K~image. This procedure produces a $``$flat-top'' profile over the center (Figure \ref{fig6}), which likely leads to a modest overestimation of the AGN contribution as the underlying stellar emission will peak at or very close to the AGN emission peak \citep[for examples of this technique, see][]{R02,R03,RA09,L09}. Then, flux of the AGN$_{\mbox{\sevensize PSF}}$-subtracted \K~image is measured to be 2.25 mJy (13.66 mag.) in a 1.2$''$ aperture. Diffuse stellar emission in the nuclear bulge has a contribution of 55$\pm$3\% (45$\pm$3\% AGN emission) as measured through this second method. 

From the methods presented above, the average of the AGN emission is estimated to be 43$\pm$3\%, whereas the contribution from the diffuse stellar emission in the nuclear bulge is estimated to be 57$\pm$3\%, both estimated in a 1.2$''$ aperture. The formal uncertainties are estimated to be 5\% from photometric measurements. Systemic errors from this methodology could increase the final uncertainties, but are difficult to quantify. Previous observations \citep{K98} at 1.1 $\um$, 1.6 $\um$ and 2.2 $\um$ using NICMOS/HST showed an unresolved point-source, consistent with an obscured AGN. \citet{K98} suggested that the infrared emission at 2.2 $\um$ is dominated by thermal emission from hot (T = 720 K) dust in the inner edge of the torus. They suggested that the 75\% of the flux at 1.6 $\um$ is due to emission other than hot dust in the inner edge of the torus, consistent with that only diffuse stellar emission contributes to the total flux at shorter wavelengths, H band.


\subsection{The nuclear intrinsic polarisation}
\label{nuclearpolflux}

The diffuse stellar emission in the nuclear regions of IC5063 significantly dilutes the observed emission from the AGN, as estimated in Section \ref{nucleartotalflux}. We subtracted the diffuse stellar emission in the measured degree of polarisation at \K~in a 1.2$''$ aperture, and  an intrinsic polarisation of P$^{\mbox{\sevensize{int}}}_{\mbox{\sevensize{\K}}} =$ 18.1$\pm$1.1\% was estimated. Note that the intrinsic polarisation calculated here is independent of the polarising mechanisms of the central engine. The estimated intrinsic polarisation is in good agreement with previous NIR studies at J, H and \K~by \citet{H87}, where an intrinsic polarisation of 17.4$\pm$1.3\% was calculated, using their own photometric data and photometry from \citet{A82} in the H and K filters using a 2.25$''$ aperture. \citet{B90b} used the same data of \citet{H87} and with slightly different assumptions calculated an intrinsic polarisation of 15.3$\pm$1.0\% in a 2.25$''$ aperture. They used a polarised power-law component and starlight subtraction. Optical spectropolarimetric studies by \citet{I93} estimated an intrinsic polarisation of $\sim$10\%, based on narrow lines in the range of 0.54 - 0.70 \um. Studies of other AGN have shown highly intrinsic polarised nucleus. For example, \citet{Simpson02} measured a polarisation of 6\% in the nucleus of NGC1068 and \citet{tadhunter2000} estimated an highly intrinsic polarisation of $>$28\% in the nucleus of Cygnus A, where both observations used the 2 $\um$ polarimetry mode of NICMOS/HST.


\section{Polarisation model}
\label{polmodel}

To investigate the polarisation from the torus, the aligned dust grains and the role of magnetic fields, we developed a polarisation model to take into account the various mechanisms of polarisation in the nuclear regions of IC5063. In the sections below, we discuss the possible mechanisms of polarisation and then we construct a polarisation model.


\subsection{Possible mechanisms of polarisation}

We consider that the nuclear polarisation could arise through three mechanisms in the NIR: (1) synchrotron radiation, as suggested by \citet{H87} from a central BL Lac type object; (2) dichroic absorption by galactic nuclear dust and the torus; and/or (3) scattering of nuclear radiation, as indicated by the ionisation cones.

\subsubsection{Synchrotron emission}

Optical and NIR polarimetric observations of IC5063 by \citet{H87} suggested that the large intrinsic degree of polarisation, 17.4$\pm$1.3\%, arises from a non-thermal nuclear source. The intrinsic polarisation is comparable with previous observations of BL Lac objects \citep{A78,BHA83} and is generally attributed to synchrotron emission mechanism. BL Lac objects are observed to show a high degree of photometric variability with times-scales of weeks and a factor of two in amplitude in the optical UBVI filters \citep*{IN88}. NIR (J, H and K bands) variation, when detect at all, are smaller and longer time-scale than in optical \citep{N89,Hunt94}. IC5063 does not show any significant flux-variability versus time \citep*{C91}, because we are using previous data for calibration, we cannot comment further on the issue of variability for IC 5063. However, the optical \citep{I93}  and NIR \citep[][this paper]{H87} polarisation shows a wavelength-dependence on the intrinsic polarisation, that is not inconsistent with a synchrotron source. A further argument against a synchrotron origin for the observed NIR polarisation is that the polarised radio emission in IC 5063 \citep{Morganti2007} is located 1$^{''}$ - 2$^{''}$ SW of the nucleus, with polarisation vectors oriented along a very different direction that what we observe in the nucleus (Figures \ref{fig1} and \ref{fig2}). A decisive test of the synchrotron emission interpretation is mm polarisation measurements, but as yet no such observations have been published. The lack of total flux variability strongly argues against synchrotron being a dominant emission mechanism. Hence, we do not consider that the polarisation arises dominantly from synchrotron mechanism. 

\subsubsection{Dichroic absorption}

\citet{Y95} showed that the polarisation at K is attributed to dichroic absorption of the central engine radiation passing through the dust within the torus with visual extinction $\geq$ 45 mag in NGC1068. They argued that the increase in the degree of polarisation with wavelength is due to dichroic absorption in the NIR. As example, \citet{Simpson02} measured a level of polarisation of 6\% in the unresolved nucleus of NGC1068 at 2$\um$ using NICMOS/HST. They argued that the polarisation is likely produced by dichroic absorption arising from the molecular clouds associated with the torus. Hence, the visual extinction of $\geq$ 45 mag. is consistent with measurements of the degree of polarisation of 6\%. For detailed discussion of the extinction to the nucleus of IC5063 at several wavelengths see Section \ref{Sec_nucext}. To study how dichroism affects the polarisation in the unresolved nucleus of IC5063, this mechanism is included in the polarisation model.

\subsubsection{Scattering}

Broad emission lines in polarised flux have been detected in the optical spectrum of IC5063  \citep{I93}, which strongly implies the presence of a nuclear dust and/or electron scattering screen. Indeed, our observations show, for first time in polarised flux in IC5063, the dual-ionisation cones at H (Figure \ref{fig5}), arising from scattering of nuclear radiation by agents within the ionisation cones. At J, the ionisation cones are not observed as (a) the counter (rear projected) cone is obscured by the dusty disc of the host galaxy; and (b) the forward (front projected) cone's polarised flux is not clearly detected due to host galaxy contamination and/or low signal to noise in the cone area. At \K, the cones are also not observed, as the signal-to-noise of our observations is insufficient to map the cones beyond the central 2$''$ aperture. To study how scattering affects the polarisation in the unresolved nucleus of IC5063, this mechanism is included in the polarising model.

To understand the polarisation mechanisms responsible for the compact nuclear polarisation and the wavelength dependencies, a polarisation model to simultaneously fit the total flux, polarised flux and the degree and PA of polarisation is developed and described in the following sections.


\begin{figure*}
\includegraphics[angle=90,scale=0.7,clip,trim=2.2cm 2cm 9cm 4cm]{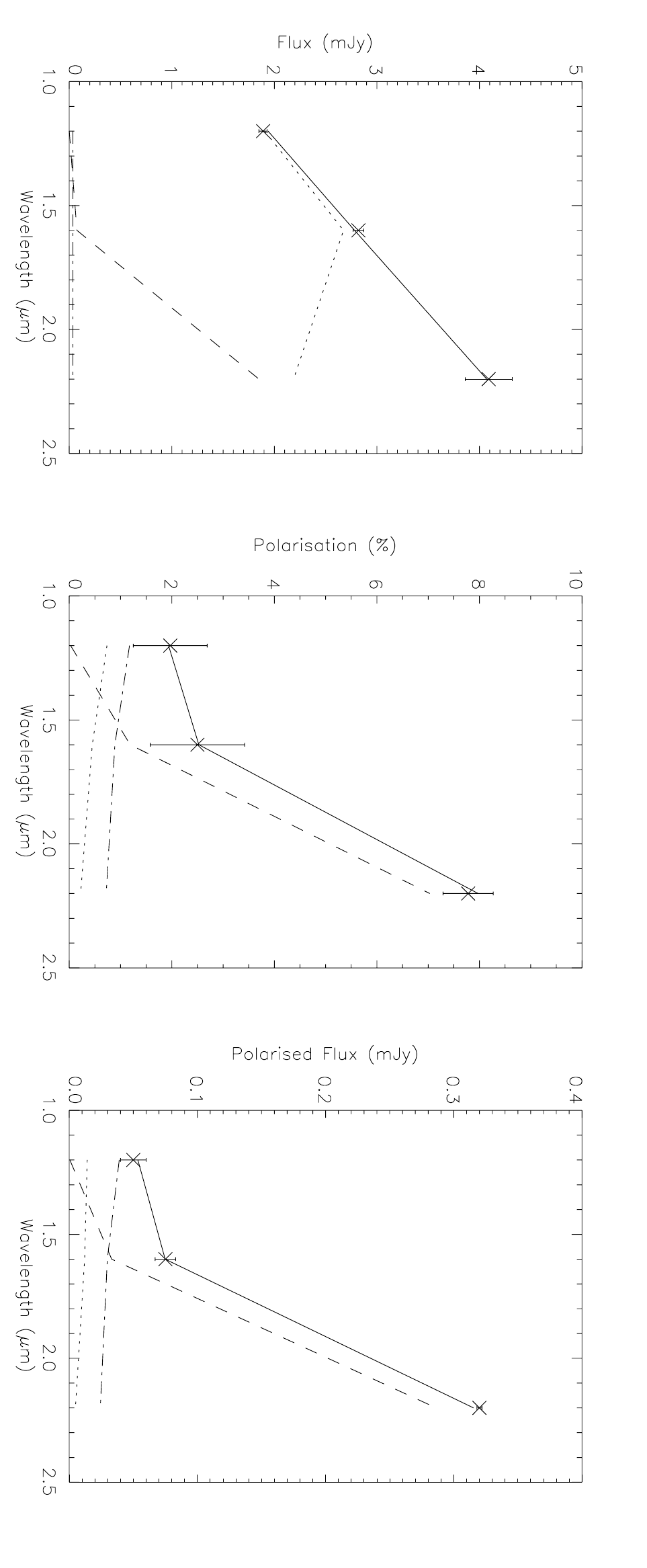}
\caption{Model fit (Section \ref{polmodel}) to the total flux in a 1.2$''$ aperture ({\it left panel}). The diffuse stellar emission through the nuclear bulge ({\it dotted line}), AGN through torus ({\it dashed line}), ionisation cones ({\it dashed dotted line}) and the total ({\it solid line}) are shown. Model fit to the degree of polarisation ({\it middle panel}) and polarised flux ({\it right panel}) in a 1.2$''$ aperture are shown. Polarisation produced through the dichroic absorption by the dust in the nuclear bulge ({\it dotted line}) and within the torus ({\it dashed line}), electron scattering ({\it dashed dotted line}) and total ({\it solid line}) are shown.} 
\label{fig7}
\end{figure*}


\subsection{Total flux modeling}
\label{fluxmodel}

To estimate the visual extinction through the torus and constrain the polarisation mechanisms in the nucleus of AGN, we developed a polarisation model to fit both the total and polarised flux in a 1.2$''$ ($\sim$ 263 pc) aperture of the type 2 AGN, IC5063.

The model consists of a point-source emitter, the central engine, partially extinguished by dust within the torus and the nuclear bulge. The central engine is described as an unpolarised power-law source, (F$_{\rm \nu} \propto \rm \nu^{-\rm \alpha}$), with an assumed power-law index, $\alpha =$ 1.5, typical of Seyfert 1 nuclei and previously used for IC5063 \citep{B90b,C91,I93}.  The power-law is partially extinguished by two contributions, one by dust in the torus, \Av(1), and one by dust in the nuclear bulge, \Av(2). We also included a contribution from the diffuse stellar emission in the nuclear bulge, modeled using typical starlight colors of elliptical galaxies, (J-H) $=$ 0.75 mag. and (H-K) = 0.22 mag. \citep{S86}. The diffuse stellar emission in the nuclear bulge contribution to the total flux was constrained to be 57\% at \K~in a 1.2$''$ aperture (Section \ref{nucleartotalflux}). The diffuse stellar emission contribution was reddened by E(B-V) $=$ 0.26 mag., estimated from the (J-H) and (H-\K) maps of IC5063. Similar values have been used in the literature, \citet{I93} used E(B-V) $=$ 0.6 mag., \citet{C91} estimated E(B-V) $=$ 0.1 - 0.4 mag., and \citet{B90b} used E(B-V) $=$ 0.12 mag., calculated by \citet*{BD86}. A range of E(B-V) $=$ 0.2-0.6 mag. was used by \citet{B83}.  The reddening power-law, $\rm \tau_{\rm \lambda} \propto \rm \lambda^{-1.85\pm0.05}$ by \citet{L84} was used.

An additional component added to the total flux to represent scattering within the NW and SE ionisation cones was also included. Both cones were (initially) assumed to suffer the same level of extinction from the nuclear bulge. 

We fit the measured total fluxes in a 1.2$''$ aperture, by adjusting the visual extinctions, \Av(1) and \Av(2), of the total flux model. The fit was considered acceptable when the deviation from the modeled J, H and \K~total fluxes was $<$1\% of the flux value at all wavelengths. Using this procedure, the fit to the total flux at J, H and \K~was obtained (Figure \ref{fig7}), giving a visual extinction of \Av(1) $=$ 48$\pm$2 mag. and \Av(2) $=$ 6$\pm$2 mag. The visual extinction for the host galaxy,  \Av(2), is close to the previous determination, \Av~$=$ 7 mag. by \citet{HR99}. The total visual extinction to the central engine of IC5063 is \Av(1) + \Av(2) $=$ 54$\pm$4 mag., consistent with the extinction, \Av~$=$ 64$\pm$15 mag., estimated using optical ([OIII]$\lambda$5007) and infrared (K and L$'$) spectral index and visual extinction maps \citep{S94}. For the ionisation cones, the SE cone (counter-cone) was extinguished with the nuclear bulge visual extinction, \Av(2) = 6$\pm$2 mag, while no extinction for the NW cone (forward-cone) was needed. This is entirely consistent with the NW cone being in front of the obscuration of the nuclear bulge.

\subsection{Polarisation modeling}
\label{polmodeling}

Using the same model, the wavelength-dependence of the observed polarised flux in the nuclear regions of IC5063 at J, H and \K~in a 1.2$''$ aperture was examined. The model assumes two separate polarising mechanisms. First, dichroic absorption by the interstellar medium of the central engine power-law emission. The interstellar polarisation follows a typical Serkowsky curve \citep*{SMF75}:
\[
\frac{\mbox{P}(\lambda)}{\mbox{P}_{\mbox{{\sevensize max}}}} = \exp \left[ -\mbox{K} \ln^{2}\left( \frac{\lambda_{\mbox{{\sevensize max}}}}{\lambda} \right)\right]
\]
\noindent
where, P$_{\mbox{\sevensize max}}$, represents the maximum of polarisation at $\rm \lambda_{\mbox{\sevensize max}}$ and K $=$ 0.01$\pm$0.05 + (1.66$\pm$0.99) $\rm \lambda_{\mbox{\sevensize max}}$ \citep{W92}. 

Our polarimetric data and the optical and NIR polarimetric data from \citet{H87} were used to fit the Serkowsky curve. The best fit is P$_{\mbox{\sevensize max}} =$ 1.4$\pm$0.1\%, $\rm \lambda_{\mbox{\sevensize max}} =$ 0.50$\pm$0.02 \um~and K $=$ 0.84$^{+0.14}_{-0.06}$. These values are similar to P$_{\mbox{\sevensize max}} =$ 1.49$\pm$0.08\% and $\lambda_{\mbox{\sevensize max}} =$ 0.51$\pm$0.02 \um, previously estimated by \citet{B90b} for IC5063.

NIR polarimetric studies have shown that the observed polarisation in the range of 1.0-2.5 $\um$ is better represented by a power law P $\propto \lambda^{-\gamma}$, with $\gamma$ in the range of 1.6-2.0 \citep[e.g.][]{Nagata1990,MW1990}. The degree of polarisation arising from dichroic absorption through a dust column can be estimated by the knowledge of the extinction, A$_{\lambda}$, and observed polarisation power-law. Thus, we use the expected polarisation by a dust column given by  P$_{\lambda}$ $\propto$ A$_{\lambda}$ $\times$  $\rm \lambda^{-\rm \gamma}$.

Second,wavelength-independent electron scattering, P $= \lambda^{0}$, is considered to be the polarising mechanism in the ionisation cones. Previous studies from UV to NIR \citep[e.g.][]{AM85, Y95,I93} have shown that electron scattering is the dominant polarising mechanism in the ionisation cones.

We fit the polarised flux, the degree and PA of polarisation using simultaneously the results of the previous section with the constraints given by (a) the intrinsic polarisation of P$^{\mbox{\sevensize{int}}}_{\mbox{\sevensize{\K}}} =$ 18.1$\pm$1.1\%; (b) the visual extinction of \Av(1) = 48$\pm$2 mag., and \Av(2) = 6$\pm$2 mag.; and (c ) a power-law index $\rm \gamma =$ 1.9 (Figure \ref{fig7}). The fit was considered acceptable when the difference between the model and observed J, H and \K~polarised flux was $<$1\%. Within the 1.2$''$ aperture, the polarised flux at J electron scattering (67\%) and dichroic absorption from dust in the nuclear bulge (33\%) are the dominant polarising mechanisms, with zero contribution of dichroic absorption within the torus. At H, the observed polarised flux is estimated to be 71\% dichroic absorption from dust in the nuclear bulge, 16\% dichroic absorption within the torus and 13\% electron scattering. At \K, the observed polarised flux produced through dichroic absorption within the torus is 89\%, 8\% electron scattering and 3\% dichroic absorption from dust in the nuclear bulge.

\subsection{Intrinsic polarisation through dichroic absorption within the torus}

The intrinsic polarisation calculated in Section \ref{nuclearpolflux} is independent of the polarising mechanisms to the central engine of IC5063. To estimate the intrinsic polarisation arising from dichroic absorption (P$^{\mbox{\sevensize{dic}}}_{\mbox{\sevensize{\K}}}$) through the torus, the measured polarisation, P$^{\mbox{\sevensize{obs}}}_{\mbox{\sevensize{\K}}} =$ 7.8$\pm$0.5\%, at \K~in a 1.2$''$ was corrected by accounting for (1) the measured degree of polarisation through dichroic absorption from dust in an off-nuclear region; (2) the estimated polarisation through dichroic absorption from dust in the nuclear bulge; (3) the fractional contribution of the central engine to the total flux; and (4) the fractional contribution of the dichroic absorption within the torus.

The degree of polarisation at \K~produced through dichroic absorption from dust in an off-nuclear region of IC5063 was measured to be P$^{\mbox{\sevensize{off}}}_{\mbox{\sevensize{\K}}} =$ 0.2\%, with a PA of polarisation $\sim$15\degree. Note that the difference in PA of polarisation at \K~between the nuclear (3$\pm$6\degree) and off-nuclear ($\sim$15\degree), represents a lower-limit in the subtraction of the diffuse stellar emission in the nuclear bulge of IC5063, as polarisation is a vector (rather than a scalar) quantity.

The estimated polarisation produced through dichroic absorption from dust in the nuclear bulge is estimated by the relationship P(\%)/\Av = 1 - 3 \% mag$^{-1}$ \citep{G78,W01,L07}. Using our estimated value of the nuclear bulge visual extinction, \Av(2) = 6$\pm$2 mag (Section \ref{fluxmodel}), transformed\footnote{The conversion of visual extinction, \Av, to extinction at \K, A$_{\rm K_n}$, is A$_{\rm K_n}$ $=$ 0.112\Av~\citep{J89}.}  to A$_{\mbox{\sevensize{\K}}}$ = 0.7$\pm$0.2 mag, the dichroic absorption polarisation from dust in the nuclear bulge at \K, is calculated to be in the range of P$^{\mbox{\sevensize{diff}}}_{\mbox{\sevensize{\K}}} =$ 0.5 - 2.7\%. 

The contributions of the AGN and diffuse stellar emission in total flux from Section \ref{nuclearpolflux} are 43$\pm$3\% and 57$\pm$3\%, respectively. We define the ratio of the diffuse stellar emission and AGN contribution to be AGN$^{\mbox{\sevensize{ratio}}} =$ 57$\pm$3\%/43$\pm$3\%. Through the use of the polarisation model described above, the fraction of polarised flux produced through dichroic absorption within the torus is AGN$^{\mbox{\sevensize{dic}}}_{\mbox{\sevensize{\K}}} =$ 89\% at \K.

The intrinsic polarisation arising from dichroic absorption (P$^{\mbox{\sevensize{dic}}}_{\mbox{\sevensize{\K}}}$) in the central 1.2$''$ aperture of IC5053 was estimated as:

\begin{equation}
\mbox{P}^{\mbox{\sevensize{dic}}}_{\mbox{\sevensize{\K}}} = (\mbox{P}^{\mbox{\sevensize{obs}}}_{\mbox{\sevensize{\K}}} - \mbox{P}^{\mbox{\sevensize{off}}}_{\mbox{\sevensize{\K}}} - \mbox{P}^{\mbox{\sevensize{diff}}}_{\mbox{\sevensize{\K}}}) \times ( 1+ \mbox{AGN}^{\mbox{\sevensize{ratio}}} ) \times \mbox{AGN}^{\mbox{\sevensize{dic}}}_{\mbox{\sevensize{\K}}} 
\end{equation}

The intrinsic polarisation at \K~produced by dichroic absorption within the torus is estimated to be P$^{\mbox{\sevensize{dic}}}_{\mbox{\sevensize{\K}}} =$  12.5 $\pm$ 2.7\%.



\section{Nuclear extinction}
\label{Sec_nucext}    
    
The AGN is obscured by the torus and suffers further extinction from the host galaxy. Both extinctions were estimated using the simple NIR polarisation model described in Section \ref{polmodel}.  However, estimates at other wavelengths provide further constraints on the extinction to the torus. For example, X-ray wavelengths offer perhaps the optimal estimation of the absorption to the central engine of the AGN, whereas the 9.7-\um~silicate feature in the MIR affords the chance to investigate the cooler dust of the torus.  In order to study the extinction at different wavelengths and its relation with the torus, the extinction to the nuclear point source was estimated through five different techniques, as described below.

\begin{enumerate}
\renewcommand{\theenumi}{(\arabic{enumi})}

\item X-ray. The visual extinction to the nucleus of IC5063 can be calculated using the standard Galactic ratio \Av/N$_{\mbox{\sevensize H}} =$ 5.23 $\times$ 10$^{-22}$ mag cm$^{2}$ \citep{BSD78}. \citet{T11} calculated an atomic hydrogen column density of N$_{\mbox{\sevensize H}} =$ 25.0$\pm$0.7 $\times$ 10$^{22}$ cm$^{-2}$ from \textit{Swift}/BAT data in the energy band of 0.5 - 200 keV. Hence, the visual extinction is estimated to be \Av~$=$ 131$\pm$4 mag. A similar result of \Av~$=$ 115$\pm$20 mag.,  was obtained by \citet{S94} using \textit{Ginga} observations. 

 \item NIR polarimetry. \citet{J89} and \citet*{JKD1992} examined the relationship between the degree of polarisation and the optical depth at K as a function of the geometry of the magnetic field. They found that  the best correlation between the degree of polarisation and the optical depth at K for stars extinguished by dust is P$_{\mbox{\sevensize K}} =$ 2.23 $\tau_{\mbox{\sevensize K}}^{3/4}$ \citep[see equation 2 in][]{J89}. This model includes with equal contributions from random and constant component of the magnetic field. For the constant component of the magnetic field, the relationship between the degree of polarisation and the optical depth is P$_{\mbox{\sevensize K}} = \tanh \tau_{\sevensize p}$ \citep[see equation A7 in][]{J89}. In the case of IC5063, using the intrinsic polarisation arising from dichroic absorption within the torus, P$^{\mbox{\sevensize{dic}}}_{\mbox{\sevensize{\K}}} =$  12.5 $\pm$ 2.7\%, and the conversion $\tau_{\mbox{\sevensize K}}$ = 0.09 \Av~\citep{J89}, the visual extinction is calculated to be in the range of 22 - 111  mag. for constant component of the magnetic field; and  equal contributions from random and constant component to the magnetic field, respectively. The visual extinction by the torus, \Av(1)$=$ 48$\pm$2 mag., estimated through the polarising model in Section \ref{polmodel}, is within the visual extinction estimated from these two relationships. 
   
\item Silicate absorption. Using the empirical relation between the visual extinction and silicate absorption strength, \Av/$\tau_{\mbox{\sevensize 9.7\um}} =$ 18.5$\pm$1.0, \citep{W87} and the silicate absorption, $\tau_{\mbox{\sevensize 9.7\um}}$ $=$ 0.32$\pm$0.02, measured from N-band spectra of the nucleus of IC5063 \citep{Y07}, the visual extinction is calculated to be  \Av~$=$ 6$\pm$1 mag. This value presumably estimates not the extinction to the central engine, but the extinction to the outer region of the clumpy torus as the observed silicate feature presumably results from a mixture of emission and absorption within the torus, this should be considered a lower limit to the true extinction.

\item Clumpy torus model. The visual extinction to our LOS, A$_{v}^{LOS}$, is calculated using the clumpy torus model \citep{N02,N08a} in IC5063. Specifically, the visual extinction produced by the torus along our LOS can be computed as \citep[see section 4.3.1 in][]{RA09}:

\begin{equation}
A_{v}^{LOS} = 1.086 N_{0} \tau_{v} e^{-((i-90)^{2}/\sigma^{2})}
\label{eq_Av}
\end{equation}

\noindent
where N$_{0}$ is the number of clouds along the equatorial direction; $\tau_{v}$ is the optical depth of the individual clouds; i is the viewing angle; and $\sigma$ is the torus angular width.

The most complete fit to IC5063 found in the literature, taking into account both photometry and spectroscopic data from NIR to MIR and a foreground dust screen geometry, is by \citet{AH11}.  We used the probabilistic distributions of the free parameters \citep[][see the blue distributions in their figure A5]{AH11} in Equation \ref{eq_Av}, to obtain a posterior probabilistic distribution of the visual extinction in our LOS \citep[for further details see][]{AR09,RA09} . Then, the median and $\pm$1$\sigma$ from the median were estimated to be $A_{v}^{LOS}$ = 1800$^{+200}_{-270}$ mag. Note that a flatter probabilistic distribution of the free parameters in the visual extinction in our LOS, A$_{v}^{LOS}$, means that a larger error is estimated. 

\item Foreground absorption. The extinction from the dust lane in the host galaxy is calculated using the (J-H) vs. (H-\K) color maps of IC5063 and its relation with visual extinction \citep{J89}. The visual extinction is measured to be \Av~$=$ 8$\pm$2 mag., consistent with \Av~$=$ 7 mag. obtained from dust features in the central 1$''$ - 2$''$ by \citet{HR99}; and  the value, \Av~= 6$\pm$2 mag., estimated using the polarisation model in Section \ref{polmodel}.

\end{enumerate} 

A summary of the various extinction values are shown in Table \ref{table4}.  The extinction at different wavelengths shows that the total flux obscuration depends strongly on the wavelength and/or model used. The estimated visual extinction through the dust-to-ratio relation in (1), is due to absorption by presumably dust free clouds close to the accretion disc, the torus and absorbers in the host galaxy. Instead, the estimated visual extinction by the torus model in (4) is due to absorption of dust in all clumps along  our LOS. Hence, the extinction calculated through these models and our polarisation model are produced by different material located in and around the torus. Based on these studies, it can be interpreted that  each wavelength penetrates different depths into the obscuring torus.


  \begin{table}
 \centering
 \begin{minipage}{85mm}
  \caption{Visual extinction at different wavelengths to the nucleus of IC5063}
  \label{table4}
  \centering
  \begin{tabular}{@{}lr@{}}
  \hline
  {\bf Method }    &    {\bf \Av }      \\
                              &  {\bf  [mag] }              \\
 \hline
     X-Ray         			   &  131$\pm$4  \\
    NIR polarisation                &  22 - 111      \\
    Polarisation Model           &   54$\pm$4                   \\
    Silicate absorption            &   6$\pm$1       \\
    Foreground absorption    &  8$\pm$2        \\
    Clumpy torus model         &  1800$^{+200}_{-270}$         \\
\hline
\end{tabular}
\end{minipage}
\end{table}

\begin{figure}
 \centering
 \begin{minipage}{85mm}
\includegraphics[angle=0,scale=0.5,clip,trim=4.2cm 8cm 6cm 5cm]{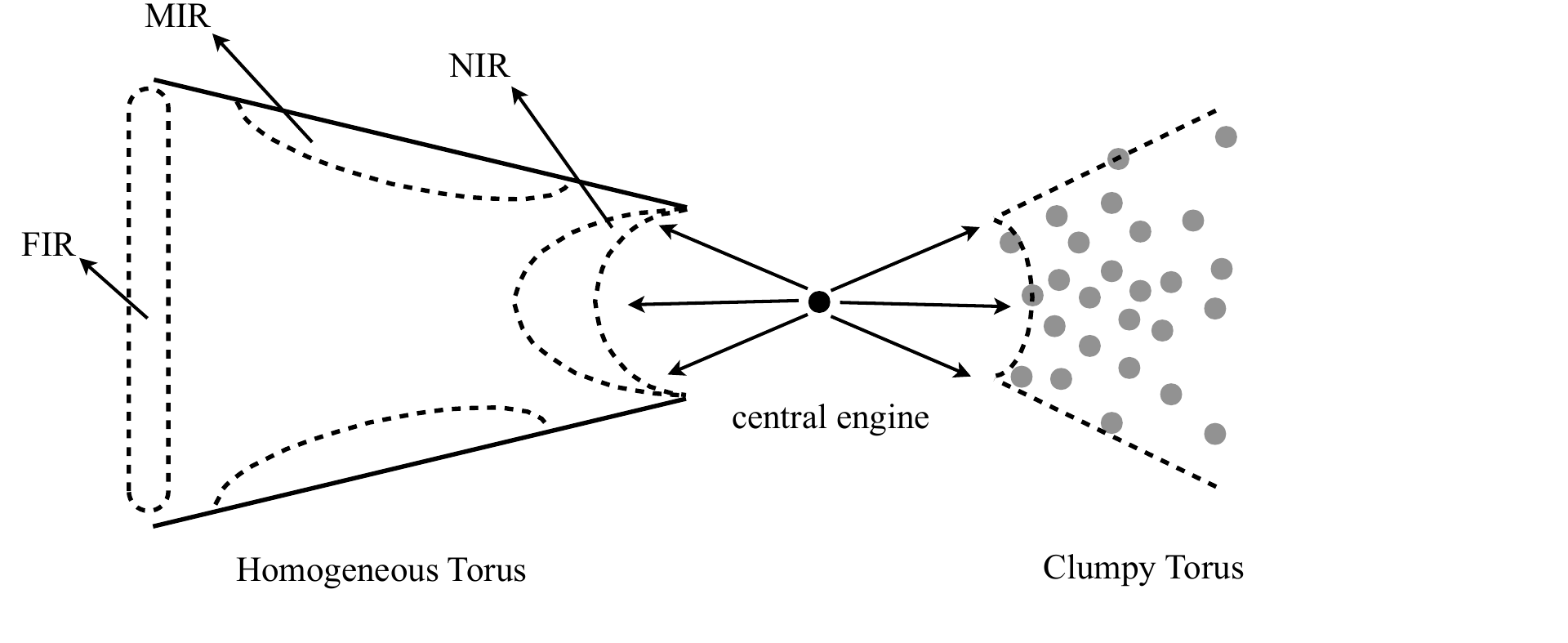}
\caption{Sketch of the homogeneous ({\it left}) and clumpy torus ({\it right}). For the homogeneous torus, the inner side of the torus emits in NIR, the middle regions emits in MIR, and the outer side emits in FIR. For the clumpy torus, the NIR emission arises from the inner side of the directly illuminated clouds, while the MIR and FIR emission is produced from the shadowed side of the clumps (see Figure \ref{fig9}). Note that clumps located at different distances can produce the same emission at a specific wavelength. See Section \ref{Sec_nucext}.} 
\label{fig8}
\end{minipage}
\end{figure}
\begin{figure}
 \centering
 \begin{minipage}{85mm}
\includegraphics[angle=0,scale=0.45,clip,trim=4.5cm 9cm 0cm 5cm]{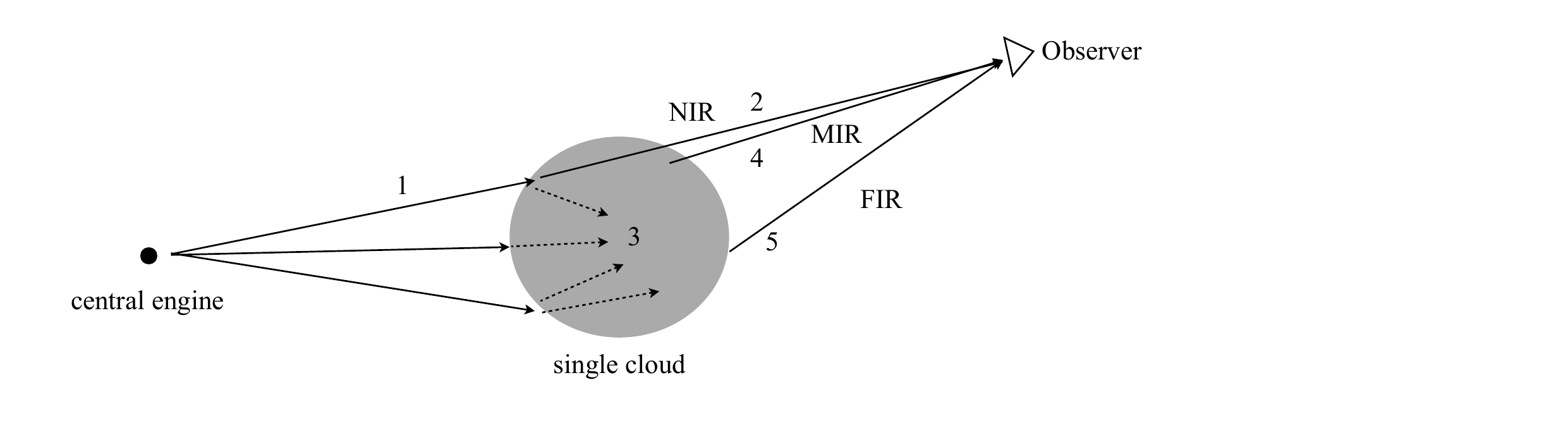}
\caption{Sketch of the emission and polarisation of an individual directly illuminated cloud in the clumpy torus. (1) The central engine emits radiation that is absorbed by the cloud, (2) the radiation in the outer layers of the cloud is polarised by the passage of light through the aligned dust grains, and the NIR polarised emission can be in the direction of our LOS. (3) Those rays with direction to the nucleus of the cloud, are completely extinguished. (4) and (5) the difference of temperature in the cloud produces that the warm dust in the middle and back side of the cloud emit in MIR and FIR, respectively. See Section \ref{Sec_nucext}.}
\label{fig9}
\end{minipage}
\end{figure}

The range of obscurations estimated are interpreted in terms of the dichroic absorption polarisation. Several NIR polarimetric studies \citep[e.g.,][]{JK89,Y95,P98,Simpson02} of AGN, have found that the nuclear polarisation of type 2 AGN typically arises through dichroic absorption within the torus to our LOS. The implications of this are (1) the total flux in the NIR wavelengths is from the directly illuminated torus inner-edge or inner-facing dust clump face (i.e. the surface of the dust clump that is directly illuminated by the central engine, Figure \ref{fig8}) or perhaps the central engine itself \citep{K05,K07,K08}; and (2) the NIR polarisation arises as the torus dust grains are aligned, most likely by the central engine's magnetic field (see Fig. \ref{fig9}).


In this scheme, an individual cloud of the clumpy torus can (dependent on the cloud's position and distribution of other clouds) absorb radiation from the central engine on the inner-facing face. This dust will re-emit radiation at NIR wavelengths in all directions.  Some of this radiation will be self-absorbed by that clump, and some will be emitted into free space.  A small amount of flux will be emitted and penetrate the dust clump at a glancing angle, leading to obscuration and dichroic polarisation from only a portion of the dust clump (see Figure \ref{fig9}).


\section{Magnetic field strength within the torus}
\label{mag_cal}

In this section the magnetic field strength in the NIR regions of the torus of IC5063 is estimated through three different methods: (1) paramagnetic alignment; (2) magnetic relaxation time; and (3) Chandrasekhar-Fermi method. Alternatively, and as a possible component of the magnetic field strength of the torus, the magnetic field strength at 1pc from the SMBH of IC5063 have been estimated (see Appendix \ref{A1}).

\subsection{Method 1: Polarisation ratio, P/\Av(\%), vs. magnetic field strength}
\label{method1}

We attributed the NIR polarisation to dichroic absorption from aligned dust grains in the clumps of the torus. The alignment can be produced by the rotational dynamics of the grain with the environment temperature and/or by the local magnetic field \citep{DG51}. The orientation of dust grains by a magnetic field is called paramagnetic alignment, through which grains are become oriented with their long axis perpendicular to the magnetic field lines.

As a first step to considering the magnetic field responsible for the aligning the dust within the torus, we consider the physical conditions and environment of the gas and dust within the torus. The gas temperature reaches a value of $\sim$ 10$^{4}$ K in the BLR \citep{N87}. NIR reverberation mapping of several AGN have shown that the outer radius of the BLR approximately corresponds to the inner radius of the dusty torus \citep{S06}. \citet{KK01} suggested that a warm absorber gas in the inner edge of the torus can reach temperatures in the range of 10$^{4}$ - 10$^{6}$ K. Recent 3D simulations of the interstellar medium surrounding the central engine showed that atomic gas and ionized [CII] trace well the inner regions of the torus, with temperatures in the range of  10$^{4}$ - 10$^{5}$ K \citep{PWS2011}. Based on these previous studies, we adopt a lower limit in the gas temperature to be T$_{gas}$ $\sim$ 10$^{4}$ K, in order to obtain a lower limit in the estimation of the magnetic field strength through methods (1) and (2). The NIR regions are located in the directly illuminated clumps in the torus from central engine\footnote{Note that several authors \citep{Mor2009,Mor2012} argued that the NIR emission can not be related with dust within the torus, but located in inner regions than the inner edge of the torus.}. \citet{N08a} estimated the dust temperatures to be in the range of 800 - 1500 K, for the directly illuminated faces of the clumps in the clumpy torus, hence we constraint the dust temperature within that range in the NIR regions studied in this paper. The temperature range is dependent on the size and grain type i.e. graphite and/or astronomical-silicates \citep{N08a}. Based on the range of temperatures, the grain size is assumed to be in the range of 0.001 - 0.01 \um. The column densities of individual clouds was calculated specifically for IC5063, assuming the following parameters: (1) \citet{AH11} obtained a radius of 2.4 pc and a number of clouds along the equatorial direction of 14$\pm$1 from the clumpy torus model; and (2) the gas column densities, as derived from NIR molecular hydrogen lines, ranging 1 - 10 $\times$ 10$^{23}$ cm$^{-2}$ \citep{D06,H09}. The column density for individual clouds in the torus of IC5063 is in the range of 10$^{4}$ - 10$^{5}$ cm$^{-3}$. A summary of these physical parameters is shown in Table \ref{table5}.


  \begin{table}
 \centering
 \begin{minipage}{85mm}
  \caption{Physical parameters assumed to the calculation of the magnetic field strength in the torus of IC5063}
  \label{table5}
  \centering
  \begin{tabular}{@{}lcr@{}}
  \hline
 {\bf Description} &  {\bf Parameter }    &    {\bf Value}      \\
   \hline
     Gas temperature    & T$_{\mbox{\sevensize gas}}$         	     &  10$^{4}$ K  \\
     Grain temperature &  T$_{\mbox{\sevensize gr}}$               &  800 - 1500 K      \\
    Grain size		 & a           							    &   10$^{-6}$ - 10$^{-5}$ cm                   \\
    Column density in the cloud &  n            				    &  10$^{4}$ - 10$^{5}$ cm$^{-3}$       \\
\hline
\end{tabular}
\end{minipage}
\end{table}


Models of paramagnetic alignment have been highly successfully applied to studies of dust grains in molecular clouds \citep[i.e.][]{L95,G95}. These studies correlate the polarisation, P(\%), with dust extinction, \Av, as a function of the magnetic field strength, B, modeled by \citet*{V81}. The efficiency of dust grains, defined as P(\%)/\Av, is directly proportional with the average alignment of the grains, and depends of the physical conditions of the environment as well as the magnetic field strength, B. An adapted version from equation 8 in their paper is presented here to be:

\begin{equation}
\mbox{P(\%)}/\mbox{A}_{\mbox{\sevensize v}} = \frac{67 \chi'' \mbox{B}^2}   {75 a \omega n}  \left( \frac{2 \pi} {\mbox{m}_{\mbox{\sevensize H}} \mbox{k} \mbox{T}_{\mbox{\sevensize gas}}} \right)^{1/2}  (\gamma - 1) \left( 1 - \frac{\mbox{T}_{\mbox{\sevensize gr}}}{\mbox{T}_{\mbox{\sevensize gas}}}  \right)
\label{P/Av}
\end{equation}
\noindent
where, $\chi''$, is the imaginary part of the complex electric susceptibility, a measure of the attenuation of the wave caused by both absorption and scattering; B is the magnetic field strength; $\gamma$ is the ratio of inertia momentum of the dust grains; T$_{\mbox{\sevensize gr}}$,  is the grain temperature; $a$, is the grain size;  $\omega$, is the orbital frequency of the grains; $n$, is the column density in the cloud; m$_{\mbox{\sevensize H}}$, is the mass of a hydrogen atom; k, is the Boltzmann constant; and T$_{\mbox{\sevensize gas}}$,  is the gas temperature.

\citet{P69} showed that the lower bound for the most interstellar grains, the ratio $\chi''$/$\omega$ is \citep[see review][]{AP73}
\begin{equation}
\frac{\chi''}{\omega} = 2 \times 10^{-12}~\mbox{T}_{\mbox{\sevensize gr}}^{-1}
\end{equation}

The ratio of the moment of inertia of the dust grains, $\gamma$, is defined as:
\begin{equation}
\gamma = \frac{1}{2} \left[ \left( \frac{b}{a} \right)^{2} + 1  \right]
\end{equation}
\noindent
where $\frac{b}{a}$ is the grain axial ratio.  A typical value of $b/a$ for interstellar dust grains is $\sim$ 0.2 \citep{AP73,KM95}.

Equation (\ref{P/Av}) was modeled for optical (V band) wavelengths by \cite*{V81}. Further studies \citep*{G95} showed that it is also applicable at K band. Using the physical conditions in Table \ref{table5}, the intrinsic polarisation arising from dichroic absorption P$^{\mbox{\sevensize{dic}}}_{\mbox{\sevensize{\K}}} =$ 12.5$\pm$2.7\% and the extinction through the torus, \Av(1) = 48$\pm$4 mag, at \K, A$_{\mbox{\sevensize \K}} =$ 5$\pm$2 mag. The magnetic field strength was estimated to be in the range of 12 - 128 mG for the NIR emitting regions of the torus of IC5063. 

We can compare our estimated magnetic field strength to previous published values. Polarimetric observations with the VLA and the GBT at 22GHz of the water vapor masers in NGC 4258, estimated the value of a toroidal magnetic field strength of 90 mG at 0.2 pc \citep{Modjaz2005}. Circular polarisation observations of NGC4258 estimated an upper-limit of the magnetic field strength in the maser features to be 300 mG \citep{H1998}. \citet{KKE1999} estimated an lower-limit of the magnetic field strength in the maser clouds of AGN to be 20 mG. These studies considered outflow winds confined in magnetic field lines generated in the central engine. Although for different objects, and at different spatial locations, our derived range of magnetic field strength compares well with these previous studies.

To estimate the magnetic field strength in the torus of IC5063, several assumptions were made. Here, we consider each of these assumptions in detail. The dust grains of the torus clumps almost certainly experience much more turbulent and extreme physical conditions than molecular clouds \citep{V81,L95,G95}, which in molecular clouds makes the alignment of dust grains less responsive to the magnetic field. 

We assumed a homogeneous magnetic field, where any inhomogeneities in the torus are ignored. This assumption has some implications for the ratio, P(\%)/\Av. If a homogeneous magnetic field is responsible for dust grain alignment in the torus clumps, then all the dust grains will be aligned along the same orientation of the magnetic field line. In this case, the ratio P(\%)/\Av~will maximize the alignment efficiency, and hence the degree of polarisation would decrease when inhomogeneities of the magnetic field are present. We interpret the magnetic field strength from Method 1, as a lower-limit to the magnetic field.

The method used here shows a strong dependence of grain sizes. Based on our interpretation of the extinction and polarising modeling, the grain sizes can be constrained. The physical conditions in the inner side of the torus (i.e. high temperature and direct radiation from the central engine) makes impossible any evolutionary grown of the grains in such region. Hence, only small grain can survive (0.001-0.1 \um). Although this physical condition allows us to refine the grains sizes ranges, the effect of the grain sizes in the above methods are difficult to quantify. 

Thus, our estimate of the magnetic field strength in the range of 12 - 128 mG represents a lower-limit for the NIR emitting regions of the torus of IC5063.

\subsection{Method 2: Magnetic relaxation time}
\label{method2}

\citet*{RDF93}  developed a computational method to solve the grain alignment problem in molecular clouds. They found that the approach followed in Method 1 (Section \ref{method1}) is valid only when the ratio of thermal to magnetic relaxation time is smaller than unity. In order to verify this condition, we calculate the lower-limit magnetic field strength required to satisfy this condition.

In the previous section, we assumed that the dust grains in the torus clumps are uniquely aligned by the presence of a magnetic field. This assumption is valid if the magnetic field is strong enough to dominate over both the turbulence within the torus clump and the rotational dynamics of the clump environment. For rotational dynamic mechanism, the dust grains are aligned when the rotational kinetic energy is coupled to the grain rotation in equilibrium with the gas temperature. The time for the dust grains to be aligned by this mechanism is given by the thermal relaxation time \citep{H88},
\begin{equation}
t_{\mbox{\sevensize thermal}} = \frac{2a \rho}{3n\mbox{m}_{\mbox{\sevensize H}}}\left(  \frac{\pi~\mbox{m}_{\mbox{\sevensize H}}}{8\mbox{k}\mbox{T}_{\mbox{\sevensize gas}}}  \right)^{1/2}
\label{t_thermal}
\end{equation}
\noindent
where, $\rho$, is the grain density.

In the case of magnetic alignment, the time for the grains  to be aligned, is given by the magnetic relaxation time \citep{H88},
\begin{equation}
 t_{\mbox{\sevensize mag}} = 1.6 \times 10^{11}\frac{a^2 \rho \mbox{T}_{\mbox{\sevensize gr}}}{\mbox{B}^2}
\label{t_mag}
\end{equation}
\noindent

Assuming paramagnetic alignment in the clouds of the clumpy torus, then the magnetic field is strong enough to align the dust faster than the rotational kinetic energy. In other words,  the magnetic relaxation time is required to be shorter than the thermal relaxation time.  i.e. t$_{\mbox{\sevensize thermal}}$ $>$  t$_{\mbox{\sevensize mag}}$. Using Equations (\ref{t_thermal}) and (\ref{t_mag}), a lower-limit of the magnetic field strength verifying that condition can be estimated:

\begin{equation}
\mbox{B}^2 > 2.4 \times 10^{11}~a~n~\mbox{m}_{\mbox{\sevensize H}} \mbox{T}_{\mbox{\sevensize gr}} \left( \frac{8\mbox{K}\mbox{T}_{\mbox{\sevensize gas}}}{\pi~\mbox{m}_{\mbox{\sevensize H}}} \right)^{1/2}
\label{B}
\end{equation}

The magnetic field strength is $>$ 2 mG and $>$ 30 mG, for the physical conditions shown in Table \ref{table5}. Note, these values are purely theoretical and assume stable physical conditions in the clouds of the clumpy torus. Hence, these values are considered as a lower-limit magnetic field strength in the clumpy torus of IC5063. Since the estimated magnetic field strength in Section \ref{method1} are larger than the lower-limit calculated here, the condition by \citet*{RDF93} is satisfied in Method 1.

Under the condition that the magnetic relaxation time is shorter than the thermal relaxation time, the gas and dust temperatures are decoupled. This assumption has two implications (1) is only valid in low-density regions in clouds; and (2) the ratio P(\%)/\Av~is dependent of the magnetic field strength. In our scheme described in Section \ref{Sec_nucext}, the detected radiation passed through the low-density regions of the torus clump, satisfying the condition described above, applied in  this section. Also, the allowance in the estimation of the magnetic field strength in Method 1.

\subsection{Method 3: Chandrasekar-Fermi method}
\label{method4}

The Chandrasekhar-Fermi method \citep*{CF53a} was also used to estimate the magnetic field strength in the torus of IC5063. This method relates the magnetic field strength with the dispersion in polarisation angles of the constant component of the magnetic field and the projection of the mean magnetic field on the plane of the sky (hereafter termed dispersion of polarisation angles, $\alpha$) and the velocity dispersion of the dust. Here (Equation \ref{CF_eq}), we use the equation 8 of \citet*{MPC2012}. This relationship is an adapted version of equation 7 in \citet{CF53a} with the factor 0.5 introduced by \citet*{OSG2001} to estimate the magnetic field strength in the plane of the sky.

\begin{equation}
B = 0.5 \left( \frac{4}{3}\pi\rho \right)^{1/2}\frac{\sigma_{v}}{\alpha}~~[\mu \mbox{G}]
\label{CF_eq}
\end{equation}

\noindent
where $\rho$ is the volume mass density in g cm$^{-3}$; $\sigma_{v}$ is the velocity dispersion in cm s$^{-1}$; and $\alpha$ is the dispersion of polarisation angles in radians. 

\begin{figure}
 \centering
 \begin{minipage}{85mm}
\includegraphics[angle=0,scale=0.8,clip,trim=0.2cm 0.2cm 0cm 0cm]{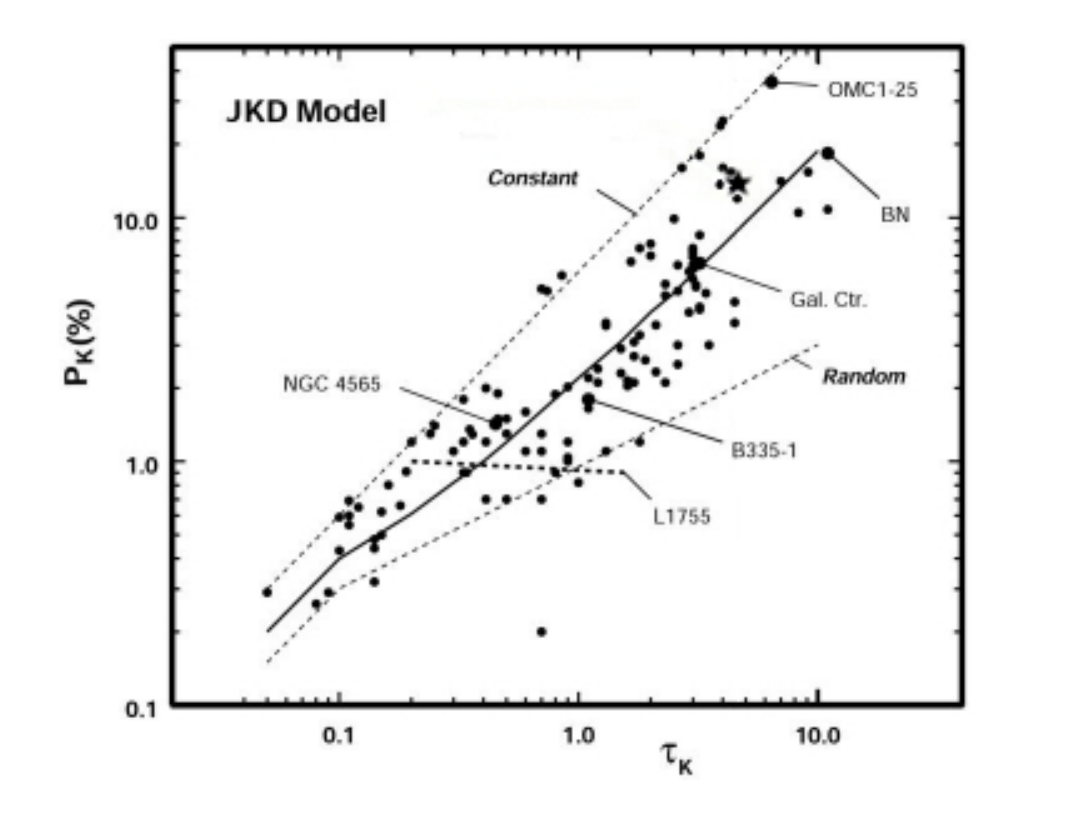}
\caption{K band degree of polarisation versus optical depth. Data from \citet{J89} ({\it black points}) with some of labeled object are shown. JKD model assuming constant, P$_{K} = \tanh \tau_p$, ({\it top dashed line}) and random, P$_{K} \propto \tau^{1/2}$, ({\it bottom dashed line}) components; and equal contributions from both constant and random components, P$_{K} = 2.23\tau_{K}^{3/4}$, ({\it solid line}) are shown.Our data, P$^{\mbox{\sevensize{dic}}}_{\mbox{\sevensize{\K}}} =$ 12.5$\pm$2.7\% and  \Av(1) = 48$\pm$2 mag. is shown ({\it star}). See Section \ref{method4}.}
\label{fig10}
\end{minipage}
\end{figure}

The volume mass density was calculated using the column mass density in Table \ref{table5} multiplied by the weight of molecular hydrogen. The velocity dispersion in masers observations in NGC 3079 is estimated to be 14 km s$^{-1}$ at a distance of $\sim$pc from the central engine \citep{Kondratko2005}. Based on the magnetohydrodynamical wind model for the torus in AGN, \citet{ES06} assumed velocity dispersion in the order of 10 km s$^{-1}$. Based on these previously published results, we use a velocity dispersion of 10 km s$^{-1}$. We note that the assumed value of the velocity dispersion represent a lower-limit. In order to calculate the dispersion of polarisation angles, $\alpha$, we used the model by \citet*[][hereafter JKD]{JKD1992}. This model relates the degree of polarisation at K with the level of turbulence in the interstellar medium and the magnetic field. In the case of IC5063, using the intrinsic polarisation arising from dichroic absorption, P$^{\mbox{\sevensize{dic}}}_{\mbox{\sevensize{\K}}} =$ 12.5$\pm$2.7\%, and the extinction by the torus, \Av(1) = 48$\pm$2 mag, we found that our data in the JKD model (Figure \ref{fig10}) is located between (1) the constant component in the magnetic field; and (2) the equal contribution of the constant and random components. If we assume that the constant component of the magnetic field is in the plane of the sky, the dispersion of polarisation angles is estimated to be $\alpha =$ 4.5\degree~(0.0785 radians) for our data point in Figure \ref{fig10}. In Equation \ref{CF_eq} we substituted the above numerical values and we estimated a lower-limit of the magnetic field strength in the plane of the sky to be 13 and 41 mG depending of the conditions within the torus of IC5063. 


We assumed that the constant component of the magnetic field strength is in the plane of the sky. If the constant component of the magnetic field is angled away from the plane of the sky, then the magnetic field strength found here will be an underestimate, for example if the magnetic field line is pointed to our LOS, we will measure zero polarisation. 

A summary of the estimated magnetic field strength through the four methods are shown in Table \ref{table6}.


  \begin{table}
 \centering
 \begin{minipage}{85mm}
  \caption{Magnetic field strength from three different methods}
  \label{table6}
  \centering
  \begin{tabular}{@{}lrr@{}}
  \hline
  {\bf Method }    &    {\bf B$_{\mbox{\sevensize min}}$}  &    {\bf B$_{\mbox{\sevensize max}}$}     \\
  			&	{\bf (mG)}		&   {\bf (mG)}	 \\
   \hline
    1: Polarisation ratio vs. magnetic field strength  							&   12	          &	128		 \\
    2: Magnetic relaxation time											&    2            	&	30	 \\
    3: Chandrasekhar-Fermi method										&   13 		&	41 	           \\
    4: Magnetic field from central engine at 1pc  								&   5 		         &	53 	           \\
    \hline
\end{tabular}
	{\bf Note:} All values represent a lower-limit of the magnetic field.
\end{minipage}
\end{table}




\section{Conclusions}

We presented NIR polarisation at J, H and \K~of the nuclear regions of IC5063. The analysis shows a highly polarised source, measured to be 7.8$\pm$0.5\% at \K, with a wavelength-independent PA of polarisation of 3$\pm$6\degree~in the three filters. For the first time in polarised light of IC5063, the biconical ionisation cones are observed, showing a spatial correspondence with the [OIII] ionisation cones and the radio structure at 8GHz, entirely consistent with unified models.

We developed a polarimetric model to account for the various mechanisms of polarisation in the central 1.2$''$ ($\sim$263 pc) aperture of IC5063. To account for the scattering pattern produced by the biconical structure observed at H, an additional polarising component due to electron scattering was required. The model of the nuclear polarisation at \K~is consistent with the polarisation being produced through dichroic absorption from aligned dust grains in the clumps of the torus with a visual extinction \Av = 48$\pm$2 mag. by the torus.

Through the use of various components to the central engine of IC5063, we estimated the intrinsic polarisation arising from dichroic absorption to be P$^{\mbox{\sevensize{dic}}}_{\mbox{\sevensize{\K}}} =$ 12.5$\pm$2.7\% at \K~in a 1.2$''$ aperture. Estimates of the extinction to the central engine of IC5063 at X-ray, NIR and MIR showed a wide variations in the extinction depending on the wavelengths on which the estimated is based on. We interpreted as that different wavelengths resulting from different emission locations within the torus and hence suffering different level of obscuration.  In this scheme, an individual cloud of the clumpy torus can, depending on the cloud position and distribution of other cloud, absorb radiation from the central engine on the inner-facing face. This dust will re-emit radiation at NIR wavelengths in all directions. Some of this radiation will be self-absorbed by that clump, and some will be emitted into free space. A small amount of flux will be emitted and penetrate the dust clump at a glancing angle, leading to obscuration and dichroic polarisation from only a portion of the dust clump.

We assumed the alignment of dust grains be produced by paramagnetic relaxation. Then, the intrinsic polarisation and visual extinction ratio, P(\%)/\Av, is a function of the magnetic field strength. We considered the physical conditions and environments of the gas and dust within the torus and we estimated the magnetic field strength in the range of 12 - 128 mG for the NIR emitting regions of the torus of IC5063. Alternatively, we estimate the magnetic field strength in the plane-of-sky using the Chandrasekhar-Fermi method. The minimum magnetic field strength in the plane-of-sky is estimated to be 13 and 41 mG depending of the conditions within the torus of IC5063. 

These studies, to our knowledge, provide the first approach investigating the magnetic field of the torus in AGN through NIR polarisation. Further NIR polarimetric observations of IC5063 and other AGN are required to refine and/or modify this approach. The next generation of polarimeters, such as adaptive optics optimized imaging polarimeter in the NIR (1-5 \um) MMT-POL \citep{Packham_MMTPOL} at the 6.5m MMT, the MIR polarimeter (7.5-13 \um) CanariCam \citep{P05} at the 10.4-m GTC  will provide a high spatial resolution and polarisation sensitivity that will allow us to refine and/or modify these studies. Also, mm-polarimetric observations with ALMA will allow to refine intrinsic properties of the torus, i.e. dust density, grain sizes, temperature, used in the methodology to calculate the magnetic fields. 



\section*{Acknowledgments}

E. Lopez-Rodriguez acknowledges support from an University of Florida Alumni Fellowship and C. Packham from NSF-0904421 grant. C. Ramos Almeida acknowledges financial support from the Spanish Ministry of Science and Innovation (MICINN) through project Consolider-Ingenio 2010 Program grant CSD2006-00070: First Science with the GTC\footnote{http://www.iac.es/consolider-ingenio-gtc/} and the Estallidos group through project PN AYA2010-21887-C04.04. A. Alonso-Herrero acknowledges support from the Spanish  Plan Nacional de Astronom\'ia y Astrof\'isica under grant AYA2009-05705-E. Supported by the Gemini Observatory, which is operated by the Association of Universities  for Research in Astronomy, Inc., on behalf of the international Gemini partnership of Argentina, Australia, Brazil, Canada, Chile, the United Kingdom, and the United States of America. E. Perlman acknowledges financial support from NSF under grant NSF-09040896. We also thank an anonymous referee for a number of helpful comments.


\bibliographystyle{mn2e}
\bibliography{References}


\appendix

\section{Magnetic field from the central engine}
\label{A1}

The central engine generates a magnetic field. The magnetic field at a given distance is proportional to the (a) magnetic field of the super massive black hole (SMBH); and (b) distance from the SMBH, following a power-law function, given by B = B$_{\mbox{\sevensize BH}} \left( \frac{\mbox{r}}{\mbox{r}_{\mbox{\sevensize BH}}} \right)^{-n}$  \citep[e.g.,][]{Silantev2009}. The magnetic field strength at the horizon event is B$_{\mbox{\sevensize BH}}$, and r$_{\mbox{\sevensize BH}}$, is the radius of the black hole horizon, which is dependent of the black hole mass, M$_{\mbox{\sevensize BH}}$. The power-law index, $n$, is taken as 5/4 from the
assumed optical thin magnetically dominated accretion disc from 5r$_{BH} < r < 100r_{BH}$ most physically significant model of \citet{Pariev2003}; and further used by \citet{Silantev2009}.  In the case of IC5063, the black hole mass, M$_{\mbox{\sevensize BH}} =$ 2.6 $\times$ 10$^7$ M$_{\mbox{\sevensize \sun}}$ was estimated by \citet{V10}, using the mass-luminosity relation and Two-Micron All-Sky Survey (2MASS) image at K-band. For black hole masses of $\sim$ 10$^7$ M$_{\mbox{\sevensize \sun}}$,  \citet{S12} estimated magnetic field strength at the event horizon in the range of B$_{\mbox{\sevensize BH}} =$ 10$^4$ - 10$^5$ G. Using these values in the above power-law function, the range of magnetic field strength is 5 - 53 mG at 1pc from the horizon event of the SMBH of IC5063.

\bsp

\label{lastpage}

\end{document}